\documentclass[preprint,12pt]{elsarticle}
\usepackage[pdftex]{hyperref}
\usepackage{graphicx}

\usepackage{amssymb}
\usepackage{geometry}
\usepackage[fleqn]{amsmath}
\usepackage{bm}
\usepackage{amsbsy}
\usepackage{lineno}
\usepackage{physics}
\usepackage{mathtools}
\usepackage{xparse}
\usepackage{accents}
\usepackage{mathtools}
\usepackage{natbib}
\biboptions{comma,round, sort}
\numberwithin{equation}{section}

\newcommand{\extk}{
\mathrm{k}
}

\newcommand{\imagi}{
\mathrm{i}
}

\newcommand{\extproduct}[4]{
\left [
\begin{array}{c}
#1 \quad #3\\
#2 \quad #4
\end{array} \right ]
}

\newcommand{\cbullet}{{\mathrlap{\bigcirc}\;\bullet}}

\ExplSyntaxOn\makeatletter
\renewcommand*\dddot[1]{%
  \placeaccent{\acc@dot\mkern1.4mu\acc@dot\mkern1.4mu\acc@dot}{#1}%
  }
\renewcommand*\ddddot[1]{%

\placeaccent{\acc@dot\mkern1.4mu\acc@dot\mkern1.4mu\acc@dot\mkern1.4mu\acc@dot}{#1}%
  }
\NewDocumentCommand \vardot {O{1} m }
  {
    \int_compare:nNnTF
      {#1} = {1}
      {\dot #2}
      {\placeaccent{\prg_replicate:nn {#1-1} {\acc@dot\mkern1.4mu}\acc@dot}{#2}}
  }
\newcommand*\placeaccent[2]{%
  \begingroup
  \def\acc@dot{\kern-0.08em.\kern-0.08em}%
  \def\acc@skip{\ifx\macc@style\displaystyle0.32
           \else\ifx\macc@style\textstyle0.32
           \else\ifx\macc@style\scriptstyle0.22
           \else0.15\fi\fi\fi ex}%
  \def\mathaccent##1##2{%
    \setbox6\hbox{$\m@th\macc@style#1$}%
    \@tempdima\wd4
    \advance\@tempdima\macc@kerna
    \advance\@tempdima-\wd6
    \divide\@tempdima\tw@
    \@tempdimb\z@
    \ifdim\@tempdima<\z@ \@tempdimb-\@tempdima \@tempdima\z@ \fi
    \vbox{\offinterlineskip
          \moveright\@tempdima\box6
          \kern\acc@skip
          \moveright\@tempdimb\box4}%
  }%
  \macc@depth\@ne
  \let\math@bgroup\@empty \let\math@egroup\macc@set@skewchar
  \mathsurround\z@ \frozen@everymath{\mathgroup\macc@group\relax}%
  \macc@set@skewchar\relax
  \let\mathaccentV\macc@nested@a
  \macc@nested@a\relax111{#2}%
  \endgroup
}
\makeatother\ExplSyntaxOff

\journal{arXiv.org}

\begin{document}

\begin{frontmatter}


\title{A new proposal to the extension of complex numbers}



\author{Israel A. Gonz\'alez Medina}

\address{Instituto Superior de Tecnologias y Ciencias Aplicadas.\\ Universidad de la Habana. \\Cuba}
\ead{israelariel.gonzalezmedina@gmail.com}

\begin{abstract}
We propose the extension of the complex numbers to be the new domain where new concepts, like negative and imaginary probabilities, can be defined. The unit of the new space is defined as the solution of the unsolvable equation in the complex domain: $|z|^2= z^* z = i$. The existence of the unsolvable equation in a closed domain as complex's lead to the definition of a new type of multiplication, for not violate the fundamental theorem of algebra.
The definition of the new space also requests the inclusion of a new mapping operation, so the absolute value of the new extended number being real and positive. We study the properties of the vector space like positive-definiteness, linearity, and conjugated symmetry. 

\end{abstract}


\end{frontmatter}

\newpage
\section{Introduction}
The accuracy of Quantum Mechanics (QM) in its realm is indisputable. QM works fine until it deals with relativistic particles, like photons that have rest mass zero and travel with a speed of $c$. Quantum field theory (QFT) is the central mathematical framework on which are developed most of the quantum mechanical models of subatomic particles to try these particles in nowadays particle physics.

The appearance of infinities is a significant problem of quantum field theory. The emergence of infinities appeared when calculations of higher orders of the perturbation series led to infinite outcomes. One way to treat these infinities is by introducing formal techniques, known as Renormalization, that modify these quantities introducing effects of self-interactions. However, the various forms of infinities suggested the divergences are related to a  more fundamental issue instead of some failures for a specific calculation. Some physicists propose alternative approaches that included changes in the basic concepts. Among them, we found the assumption of the existence of negative or incomplete probabilities.

Paul Dirac introduced the concept of negative energies and negative probabilities, in 1942, when he quoted in the article "The Physical Interpretation of Quantum Mechanics" \cite{DiracInterpQM} that: "Negative energies and probabilities should not be considered as nonsense. They are well-defined concepts mathematically, like a negative of money." Also, Richard Feynman argued on its work "Negative probabilities" \cite{Feynman1987FEYNP} how not only negative probabilities but probabilities different from unity could be useful in probability calculations. Negative or incomplete probabilities have later been proposed to solve several problems and paradoxes.

Although there is no mathematical reason for not defining the concepts of negative and incomplete probability, their physical interpretation is as complicated as diverse. In quantum mechanics, the probability for a state being measured is the modulus squared of the wave function, which is a complex quantity and is usually represented like $\psi$. Then, mathematically speaking, the assumption of a negative probability can be written as the existence in the theory of an expression like $\psi^* \psi < 0$. The inclusion of the quantum concept of the negative probability implies the mathematical need for extending the complex numbers for equations like $z^* z <0 $ can take place.

On the other side, this work is a correction of the similar chapter of a more general research entitled \href{http://arxiv.org/abs/1811.12175}{A new proposal for a quantum theory for isolated n-particle systems with variable masses connected by a field with variable form} \cite{Israel:1811.12175}. The first three chapters of the work revise the classical mechanic for systems with variable masses connected by a field with variable form ($n$-VMVF systems) and obtain two sets of Hamilton equations for solving the problem. From the Hamilton theory, two canonical transformations are needed to evolve the system: one using the rectangular coordinates and another using the angulars. The bi-dimensional canonical transformation of the canonical variables of the system points out that the quantum theory for $n$-VMVF systems might be represented with two-component vectors, different than the complex Hilbert space. This new space can be defined from the expansion of the complex vector space. 

The classical result, together with the possibility of including the negative probability, is the main motivation and guide for defining the new space and expanding the complex numbers.

In the second section, we briefly recall the obtaining process and the history of the complex numbers as the starting point for their expansion to the new domain. Next, in section 3, we define the new domain's unit and apply the vector space's axioms to the new proposed inner product, which reveals the need to include new operations to the extended domain. In section 4, we formally define the new operations and their properties. They are the bases, in section 5, to analyze the algebraic properties of extended numbers like Associativity, Commutativity, Distributivity, and the Existence of identity and elements. In section 6, we define and study the new domain's inner product. In section 7, we study the division between two numbers on the new domain by reviewing extraneous and missing solutions when the same factor multiplies both members of an equation. In sections 8, 9, and 10, we study some of the properties of the inner product and its Linearity and Conjugate symmetry. In section 11, we summarize the method and the equations for computing the needed equations to define the new space. Finally, we present the conclusions in chapter 12.
 
\section{History}

Numbers are fundamentals in science. They are mathematical objects used to describe quantity, order, or measure of concepts like time, space, matter, and fields. Physics theories like classical mechanics, electromagnetism, and quantum theory are developed using analytic entities like metrics space, tensors, fields, whose operations between them are based on the properties of the numbers that the theory lays on. Numbers are classified as $i)$ natural, $ii)$ integer, $iv)$ rational, $v)$ real, and $vi)$ complex. The Complex numbers are defined as the solution of unsolved equations in the real domain, such as $x^2 + 1 = 0$, from where the complex unit is defined as $\imagi = \sqrt{-1}$. 

The fundamental theorem of algebra states that every non-constant single-variable polynomial with complex coefficients has at least one complex root, or from the algebraic point of view, the field of complex numbers is algebraically closed for sum and multiplication operations. Thus, the search for extending the given above classification to numbers of high dimensionality may seem fruitless. However, based on advanced concepts of modern algebra, number systems called quaternions, tessarines, coquaternions, biquaternions, and octonions were developed in the nineteenth-century as extensions of complex numbers. They are all covered by the concept of hypercomplex number. 

The study of hypercomplex numbers began in 1872 when Benjamin Peirce and, later, his son Charles Sanders Peirce published the Linear Associative Algebra \cite{10.2307/2369153}. They identified the nilpotent and the idempotent elements to classify the hypercomplex numbers. Later, involutions were used in the Cayley–Dickson construction to generate complex numbers, quaternions, and octonions.  Notorious theorems were proved about the limit of hyper-complexity such as Hurwitz (normed division algebras) and Frobenius's theorems (associative division algebras). Also, in 1958 J. Frank Adams prove that there exist only four finite-dimensional real division algebras: the reals $\mathbb{R}$, the complexes $\mathbb{C}$, the quaternions $\mathbb{Q}$, and the octonions $\mathbb{O}$ \cite{10.2307/1970147}. Later Clifford develops what is known in the literature as Clifford algebra that generalizes the real numbers, complex numbers, quaternions, and several other hypercomplex number systems and is connected with the theory of quadratic forms and orthogonal transformations.

One of the most famous is the quaternions, which were first described by William Rowan Hamilton in 1843 as the quotient of two directed lines in a three-dimensional space or equivalently as the quotient of two vectors \cite{hamilton1866elements}. A quaternion is usually represented as
\begin{equation}
a + b\mathbf{i} + c\mathbf{j} + d\mathbf{k}
\end{equation}
where $a, b, c$, and $d$ are real numbers, and $\mathbf{i}, \mathbf{j}$, and $\mathbf{k}$ are the fundamental quaternion units. A notable property of a quaternion is that the multiplication of two quaternions is noncommutative.

Different from the work developed by Hamilton and others great mathematicians that develop the previous number systems, the objective of the present article is to propose a new number system that includes the arithmetical concept of the negative probability that can be used for the construction of the new Lorentz space on which can be developed a new proposal for the quantum mechanic for the $n$-VMVF systems.

\section{The complex extended unit $\extk$}

We construct the new domain in the same way complex number was defined: as the solution of an unsolvable equation in the domain about to be extended. The chosen unsolvable equation on the domain of complex numbers $\mathbb{C}$ is
\begin{equation}
|z|^2 = \imagi.
\end{equation}
We can name the new set of numbers as the Extended Complex, because of the inclusion of the conjugated complex numbers, and we can represent it by $\mathbb{E}$. The unit for this new set of numbers can be defined as
\begin{equation}
|\extk|^2\equiv \extk^* \extk = \imagi \qquad \text{ or } \qquad \extk = \sqrt[*]{\imagi} \qquad \extk \in \mathbb{E} \label{extConjProdDef}
\end{equation}
being $\mathbb{E}$, the set of the extended complex numbers and $\sqrt[*]{()}$ is the conjugated square root of a number $x$. The conjugated square root is the operation that results in a number $y$ such that its product over its complex conjugate is  $x$: $y^*y=x$. 

It is very important to remark that this unsolvable equation is not unique, and the extended unit's definition may be different. We can set a general unsolvable equation for defining the extended unit as
\begin{equation}
|z|^2 = \exp^{\imagi \theta} \qquad \forall\; 0 < \theta < 2\pi.
\end{equation}
The definition of the extended unit in this proposal is established as above, $\extk = \sqrt[*]{\imagi}$. However, this expression must be considered as replaceable in case of future improvements needed.

We can express a general extended complex number as
\begin{equation}
 \alpha = x\extk + y   \qquad \qquad \forall x,y \in \mathbb{C}.
\end{equation}
or replacing the complex numbers with their real components:
\begin{equation}
\alpha=a  \imagi  \extk + b \extk + c \imagi + d, \qquad \forall a,b,c,d \in \mathbb{R}.
\end{equation}

As the absolute value of the imaginary unit $\imagi^* \imagi = 1$ and because of our extended unit definition, $\extk^* \extk = \imagi$, we lost no generality by imposing that the absolute value for the extended unit must be equal to $1$. It implies the necessity to define a new map, besides the conjugated map, for the absolute value of the extended number being real. The new map, $ \mathcal{O}_{\text{new}}(\extk)$, appears with the definition of the new domain. So when acting on a pure complex number, it must keep the number invariant, the same way when the conjugated map, $()^*$, acts over real numbers. Then
\begin{equation}
\mathcal{O}_{\text{new}}(x) = x \qquad \forall \quad x \in \mathbb{C}.
\end{equation}

If this new map, $\mathcal{O}_{\text{new}}(x)$, is proven as a homomorphism, or the inverse of the previous assumption was satisfied, at least for the pure complex numbers, then an absolute value is a product that should have at least four factors. Indeed, if we proposed the absolute value for the extended unit as
\begin{equation}
1 = \mathcal{O}_{\text{new}}(k) \extk^* \extk = \mathcal{O}_{\text{new}}(k) \imagi \qquad \text{then} \qquad \mathcal{O}_{\text{new}}(k)=\imagi^*,
\end{equation}
which, according to the invariant property of the new map over complex numbers, led to $k=\imagi^*$. For the absolute value of the extended unit, we propose then a four term's expression like
\begin{align}
\mathcal{O}_{\text{new}}(\extk^*) \mathcal{O}_{\text{new}}(\extk)=\imagi^* \quad \text{so}\quad \mathcal{O}_{\text{new}}(\extk^*) \mathcal{O}_{\text{new}}(\extk)\;\;\extk^*\extk= \imagi^*\imagi=1
\end{align}
or
\begin{align}
\mathcal{O}_{\text{new}}^*(\extk) \mathcal{O}_{\text{new}}(\extk)=\imagi^* 
\quad \text{so}\quad 
\mathcal{O}_{\text{new}}^*(\extk) \mathcal{O}_{\text{new}}(\extk)\;\;\extk^*\extk= \imagi^*\imagi=1,
\end{align}
where we have subtly use the associativity property of the multiplication between pairs  $\mathcal{O}_{\text{new}}^*(\extk) \mathcal{O}_{\text{new}}(\extk) $ or $\mathcal{O}_{\text{new}}^*(\extk) \mathcal{O}_{\text{new}}(\extk)$ and $\extk^*\extk$.

We represent this operation $\mathcal{O}_{\text{new}}(\extk)$ as $\extk^\bullet$, so we have the first definition:
\begin{align}
&\extk^* \extk =\imagi
\nonumber \\
&(\extk^*)^\bullet \extk^\bullet = \imagi^* \label{extUnitDef}
\end{align}
or
\begin{align}
&\extk^* \extk =\imagi
\nonumber \\
&(\extk^\bullet)^* \extk^\bullet = \imagi^*. \label{extUnitDef1}
\end{align}

The operations $\extk^*$ and $\extk^\bullet$ should transform one point of the space into another point from the same space. The operations should have the general form:
\begin{align}
\extk^*= z_1 \extk+w_1
\nonumber \\
\extk^\bullet = z_2 \extk+w_2 \label{extMappProp}
\end{align}
where $z_i$ and $w_i$ are complex numbers to be determined.

However, this assumption and the definition of the extended unit lead to some inconsistencies. For example, from that definition, and replacing $\extk^* = z_1 \extk + w_1$, we obtain 
\begin{equation}
 \extk^* \extk = (z_1 \extk+w_1)\extk =  \imagi \to \extk^2 = -\frac{w_1}{z_1} \extk + \frac{\imagi}{z_1}.
\end{equation}
This expression constraints operation $\extk^2$, contradicting the independence that should exist between such operations. If $\extk^2$ and $\extk^* \extk$ are related, then the unsolvable equation from where the extended unit was defined can be related to an expression containing $\extk^2$. In that case, we will not have any unsolvable equation, because $\extk^2$ is defined at all the complex domain. From the second equation of any of the expressions \ref{extUnitDef} or \ref{extUnitDef1}, we can extract another independent equation for $\extk^2$. It relates the complex numbers $z_1$, $z_2$, $w_1$, $w_2$, and with them, the operations $\extk^*$ and $\extk^\bullet$, which should be independent between them.

\section{Introduction to the extended numbers}
The definition of the extended unit as $\extk^*\extk = \imagi$ might seem contradictory since the fundamental theorem of algebra states that the set of complex numbers is algebraically closed for the sum and multiplication operations, so it should not exist an unsolvable equation for an $n$-degree polynomial. Well, there are not. The unsolvable equation on a closed space only appears if we include the complex conjugated numbers in the polynomial. Indeed, from a polynomial like
\begin{equation}
a_n x^n + a_{n-1} x^{n-1} + ... + a_2 x^2 + a_1 x + a_0 = \sum_n a_n x^n=0 \qquad \forall \; a_n \in \mathbb{R}
\end{equation}
could never be extracted from the unsolvable equation used for defining the extended unit $\extk$. Instead, an equation like $|x|^2 = x^* x = i$ can be extracted from a different type of polynomial, such as:
\begin{equation}
\sum_{n, n'} a_{n,n'} (x^*)^n x^{n'}=0 \qquad \forall  \; a_{n,n'} \in \mathbb{R}. \label{complexPoln}
\end{equation}
This kind of polynomials that includes the conjugated numbers will have unsolvable equations like the one using for the new unit's definition, signalizing the possibility of expansion without violating the fundamental theorem in algebra. 

In Abstract, an algebraic structure is defined as the junction of a nonempty set \textbf{\textit{A}}, a collection of operations on \textbf{\textit{A}} of finite arity (typically binary operations) and a finite set of identities, known as axioms, that the set and the operations must satisfy. For example, the ring of the complex numbers is algebraically represented as
\begin{equation}
(\mathbb{C};+,\cdot).
\end{equation}
The inclusion of conjugated complex numbers in the above polynomials \ref{complexPoln} can be done by adding those numbers to the set of scalars or adding new mathematical operations that resemble the existing operations and include the conjugated complex numbers. The operations are the sum and multiplication with the complex numbers. The algebraic structures representing previous cases are
\begin{equation}
(\mathbb{E} \cup \mathbb{E}^*;+,\cdot) \label{extNumberStruc0}
\end{equation}
and 
\begin{equation}
(\mathbb{E};+,\cdot,\oplus, \odot) \label{extNumberStruc}
\end{equation}
respectively. The development of any case should be in correspondence with the other. On this work, we study the new algebraic structure that adds two new operations to the structure as the last case \ref{extNumberStruc}.

We introduced the new sum operation of two complex numbers $a,b$, as the conjugated sum. The operation is composed of two steps: $i)$ to transform the first added into its complex conjugated and $ii)$ sum the result with the second number like
\begin{equation}
a\oplus b \equiv a^* + b.
\end{equation}
The new product of two complex numbers is designed as the conjugated product and is explicitly written like
\begin{equation}
a\odot b \equiv a^* b .
\end{equation}
The operation is composed of two steps: $i)$ to transform the first number into its complex conjugated and $ii)$ multiply the result with the second number.

From here and below, we must consider the absence of an operator in a product of two extended numbers, $\alpha \beta$, as an erroneous expression, since the rules for the multiplication of the extended unit are undetermined. Nevertheless, the multiplication with no explicit operator is allowed once the rules are already defined and applied.

The expansion of the conjugated product of two extended numbers that can be written as
\begin{align}
\alpha \odot \beta = \alpha^* \beta =(x \extk + y)^*(u \extk + v) = (x^* \extk^* + y^*)(u \extk + v),
\end{align}
includes the expression $\extk^* \extk$. As a general form, we consider that the conjugated mapping depends on the number, $\alpha = x\extk + y$, which is applied on. However, the extended unit's definition, and with it, $\extk^*  \extk= \imagi $, should remain for all the extended numbers. That means that the extended unit's definition needs to be replaced with the more general expression:
\begin{equation}
\extk^*  \extk \to \extk^{(\alpha)} \odot \extk = \extk^{*(\alpha)}  \extk= \imagi \label{extConjProdDef2}
\end{equation}
and conjugated product of two extended numbers is written as
\begin{align}
\alpha \odot \beta &= \alpha^* \beta =(x \extk + y)^*(u \extk + v) 
\nonumber \\
&= (x^* \extk^{*(\alpha)} + y^*)(u \extk + v)
\nonumber \\
&= x^*u \extk^{*(\alpha)} \extk + x^*v \extk^{*(\alpha)}  +  y^*u \extk +  y^*v.
\end{align}
If $\extk^{*(\alpha)} = z_1^{(\alpha)} \extk + w_1^{(\alpha)}$ and equation \ref{extConjProdDef2}, the term $\extk^{*(\alpha)}  \extk$ leads to the relation
\begin{equation}
\extk^{*(\alpha)}  \extk= \imagi \qquad \to \qquad \extk^2 = -\frac{w_1^{(\alpha)}}{z_1^{(\alpha)}} \extk + \frac{\imagi}{z_1^{(\alpha)}} \label{extConjProdDef1}
\end{equation}
where $z_1^{(\alpha)}, w_1^{(\alpha)}$ corresponds to the extended and the imaginary parts of the complex conjugated map of the extended number.

On the other side, the standard product of the extended unit $\extk \cdot \extk$ must also remain invariant for all sets of extended numbers. Different from the definition we gave to the extended unit, the square operation does not arise from an unsolvable equation. Nevertheless, it must be defined for all extended numbers. We define the standard operation between the extended unit $\extk \cdot \extk$ as:
\begin{equation}
\extk \cdot \extk = \extk^2 = z_0 \extk + w_0 \qquad \forall \;\mathbb{E}. \label{extStandProdDef}
\end{equation}
where $z_0, w_0$ are two complex numbers to be defined. The definition set an independent expression of $\extk^2$ from \ref{extConjProdDef1}. The standard multiplication of the extended unit, as on complex numbers, does not depend on the product's factors. Hence the constant character of $z_0, w_0$  and the $\forall \;\mathbb{E}$. Using this definition, the expression 
\begin{equation}
\extk^{*(\alpha)} \cdot \extk  = \extk^{*(\alpha)} \extk = (z_1^{(\alpha)} \extk + w_1^{(\alpha)})\extk = (z_0z_1^{(\alpha)} + w_1^{(\alpha)})\extk + z_1^{(\alpha)} w_0 \neq \imagi \label{extStandProdDef1}
\end{equation}

Resuming, within the extended numbers, we define not one product but two: the standard and the conjugated product. They are just two common products following different rules for the expressions $\extk^{*(\alpha)}\odot  \extk $ and $\extk \cdot  \extk $. The definitions that define those rules are
\begin{equation}
\extk^{*(\alpha)} \odot  \extk = \imagi \qquad \text{ and } \qquad \extk\cdot \extk = z_0 \extk + w_0. \nonumber
\end{equation}
To compute these operations, we must compute the complex extended for the first factor if needed (only for the conjugated product), then establish the rules to be used for the expressions involving the extended unit, $\extk^{*(\alpha)} \extk$, and $\extk \cdot \extk = \extk^2$, and finally apply them to the result.

Being $\extk^{*(\alpha)} = z_1^{(\alpha)} \extk + w_1^{(\alpha)}$, the expressions $\extk^2$ and $\extk^{*(\alpha)} \extk$, for the standard product must compute as equations \ref{extStandProdDef} and \ref{extStandProdDef1}, while for the conjugated product the expression have the form like reference \ref{extConjProdDef1}. The table \ref{extDefTable} resumes the different expressions for every case.
\begin{table}[h]
\caption{Expressions for $\extk^2$ and $\extk^{*(\alpha)} \extk$ for each extended product} \label{extDefTable}
\begin{center}
\begin{tabular}{@{} c  c  c @{}}
\hline \hline
 & $\alpha \cdot \beta$ & $\alpha \odot \beta$	\\ 
\hline
$\extk^2$ & $z_0 \extk + w_0 $ & $ -\frac{w_1^{(\alpha)}}{z_1^{(\alpha)}} \extk + \frac{\imagi}{z_1^{(\alpha)}} $  \\

$ \extk^{*(\alpha)}\extk$ &$  (z_0z_1^{(\alpha)} + w_1^{(\alpha)})\extk + z_1^{(\alpha)} w_0$ & $  \imagi$ \\
\hline
\end{tabular} 
\end{center}
\end{table}

As $\oplus$ and $\odot$ are a type of sum and multiplication operations, the order of priority of the operations then is similar. That means that any multiplication is granted higher precedence than any addition. However, the order of priority between standard and conjugated operations, multiplication or addition, should be emphasized with parentheses $(\;)$ or brackets $[\;]$. 

The choice of expressing an extended number like $x \extk + y$, where $x$ and $y$ are complex numbers and the standard product with the expression $\extk \cdot \extk = z_0 \extk + w_0$, responds to the preposition that state that ``the Cartesian product of two rings is also a ring'' as shown on chapter 2 two of reference \cite{DaSilvaSouza2009:thesis.chapter2}. From this preposition we can conclude that the set $\mathbb{C} \times \mathbb{C} = \mathbb{C}^2$ is a ring with the operations:
\begin{align}
(a,b) + (c,d) = (a + c, b + d) \nonumber \\
(a,b) + (c,d) = (ac, bd).
\end{align}
As shown below, the extended numbers set with the standard sum and multiplication also behave as rings.

Lets now show both types of extended products between two extended numbers using both rules:
\begin{itemize}
\item The standard product of two extended is pretty straightforward, computing the product as usual and replacing \ref{extStandProdDef}. For the extended numbers $\alpha =x \extk + y$ and $\beta = u \extk + v$:
\begin{align}
\alpha \cdot \beta &=(x \extk + y)\cdot (u \extk + v) = xu \extk^2 + (xv + yu)\extk + yv
\nonumber \\
&=  xu (z_0 \extk + w_0) + (xv + yu)\extk + yv
\nonumber \\
&=  (xu z_0 + xv + yu)\extk + xuw_0+ yv.
\end{align}

Let us verify that computation of the standard product is independent of choosing $\extk^*$ or $\extk^2$ expression. First, we express a complex extended number as the complex conjugate of the conjugated square root:
\begin{equation}
\alpha  = (\sqrt[*]{\alpha})^* \equiv \alpha^{'*}.
\end{equation}
We can relate their coefficients as:
\begin{align}
x \extk + y  &= (x'\extk + y' )^* \nonumber \\
&= x^{'*} z_1^{(\alpha')} \extk + x^{'*} w_1^{(\alpha')} + y', 
\end{align}
from where
\begin{equation}
x' = \frac{x^*}{z_1^{*(\alpha')}} \qquad \text{and} \qquad y' = y^* - \frac{x^*}{z_1^{*(\alpha')}} w_1^{*(\alpha')}. \label{complexRootcCoeffRelations}
\end{equation}

The standard product of $\alpha^{'*}$ and $\beta$ can be found using any of the expressions of the table \ref{extDefTable}. We replace the map $\alpha^{'*}$ and then we proceed with the standard product:
\begin{align}
\alpha^{'*} \cdot \beta =(x' \extk + y')^*\cdot (u \extk + v) = (x^{'*} \extk^{*(\alpha)} + y^{'*})\cdot (u \extk + v) \label{standardProductDef0}
\end{align}
then we replace $\extk^{*(\alpha')}$ and compute the standard product in the usual way with \\ $\extk^2 = z_0 \extk + w_0$
\begin{align}
&(x^{'*} \extk^{*(\alpha')} + y^{'*})\cdot (u \extk + v) = [x^{'*} ( z_1^{(\alpha')}\extk + w_1^{(\alpha')})+ y^{'*}]\cdot (u \extk + v)
\nonumber \\
 = &x^{'*} u z_1^{(\alpha')}\extk^2 + ( x^{'*} v z_1^{(\alpha')} + x^{'*} u w_1^{(\alpha')} + y^{'*}u)\extk + x^{'*} v w_1^{(\alpha')} + y^{'*}v
\nonumber \\
= &( x^{'*} u z_0 z_1^{(\alpha')} + x^{'*} v z_1^{(\alpha')} + x^{'*} u w_1^{(\alpha')} + y^{'*}u)\extk + x^{'*} u w_0 z_1^{(\alpha')} + x^{'*} v w_1^{(\alpha')} +  y^{'*}v. \label{standardProductDef1}
\end{align}
Replacing equations \ref{complexRootcCoeffRelations}, we obtain
\begin{align}
\alpha^{'*} \cdot \beta &= ( x u z_0  + x v  + x u \frac{w_1^{(\alpha')}}{z_1^{(\alpha')}} + uy - x u \frac{w_1^{(\alpha')}}{z_1^{(\alpha')}})\extk 
+ x u w_0  + x v \frac{w_1^{(\alpha')}}{z_1^{(\alpha')}}  +  yv - x v \frac{w_1^{(\alpha')}}{z_1^{(\alpha')}} \nonumber \\
&= ( x u z_0  + x v  + uy )\extk + x u w_0   +  yv 
\end{align}
which is the same result as the standard product $\alpha \cdot \beta$. Now, we compute the same product $\alpha^{'*} \cdot \beta$, replacing this time the expression $\extk^{*(\alpha')} \extk$ in equation \ref{standardProductDef0} with
\begin{equation*}
\extk^{*(\alpha')} \extk = (z_0z_1^{(\alpha')} + w_1^{(\alpha')})\extk + z_1^{(\alpha)'} w_0.
\end{equation*}
In that case, we have:
\begin{align}
&(x^{'*} \extk^{*(\alpha')} + y^{'*})\cdot (u \extk + v) = x^{'*} u \extk^{*(\alpha')} k +  x^{'*}v \extk^{*(\alpha')} + y^{'*} u \extk + y^{'*} v
\nonumber \\
&=( x^{'*} u z_0 z_1^{(\alpha')} + x^{'*} v z_1^{(\alpha')} + x^{'*} u w_1^{(\alpha')} + y^{'*}u)\extk + x^{'*} u w_0 z_1^{(\alpha)'} + x^{'*} v w_1^{(\alpha')}  + y^{'*}v,
\end{align}
which is the expression as result \ref{standardProductDef1}, showing the consistency of the two rules for the standard product.

\item The conjugate product can be computed spanning the product using the distributive law, applying the complex conjugated map, and using the definition $\extk^{*(\alpha)} \extk =\imagi$. For the extended numbers $\alpha=x \extk + y$ and $\beta = u \extk + v$ we have
\begin{align}
\alpha \odot \beta &= (x \extk + y)\odot(u \extk + v) = x^*u \extk^{*(\alpha)} \extk + x^*v\extk^{*(\alpha)} + y^*u\extk + y^*v
\nonumber \\
&=   x^*u \imagi + x^*v(z_1^{(\alpha)} \extk + w_1^{(\alpha)}) + y^*u\extk + y^*v
\nonumber \\
&=   (x^*vz_1^{(\alpha)}  +  y^*u )\extk + x^*u \imagi +x^*vw_1^{(\alpha)}+ y^*v.
\end{align}

The expression can also be computed, applying first the complex conjugated map and then the distributive law. In this case, the quantity $\extk^2$ must have the definitions according to table \ref{extDefTable}:
\begin{equation}
\extk^2 = -\frac{w_1^{(\alpha)}}{z_1^{(\alpha)}} \extk + \frac{\imagi}{z_1^{(\alpha)} }.
\end{equation}
The conjugate product then have the expression
\begin{align}
&(x \extk + y)\odot(u \extk + v) =(x^* \extk^* + y^*)(u \extk + v) 
\nonumber \\
&=(x^*z_1^{(\alpha)} \extk + x^*w_1^{(\alpha)} +  y^*)(u \extk + v)  
\nonumber \\
&= x^*u z_1^{(\alpha)}\extk^2 + (x^*v z_1^{(\alpha)} + x^*uw_1^{(\alpha)} +  y^*u)\extk + x^*vw_1^{(\alpha)} +  y^*v
\nonumber \\
&= (x^*v z_1^{(\alpha)} + x^*uw_1^{(\alpha)} +  y^*u - x^*uw_1^{(\alpha)})\extk + x^*vw_1^{(\alpha)} +  y^*v +  x^*u \imagi
\nonumber \\
&=   (x^*vz_1^{(\alpha)}  +  y^*u )\extk + x^*u \imagi +x^*vw_1^{(\alpha)}+ y^*v ,
\end{align}
which match the result computed with rule $\extk^{*(\alpha)} \extk =\imagi$.
\end{itemize}

The standard and the conjugated sum of two extended number are
\begin{align}
\alpha + \beta = (x \extk + y) + (x \extk + y) &= (x + u)\extk +( y+ v)
\end{align}
and
\begin{align}
\alpha \oplus \beta &=  (x \extk + y) \oplus (u \extk + v) 
\nonumber \\
&= (x \extk + y)^* + (u \extk + v) = (x^* \extk^{*(\alpha)} + y^*) + (u \extk + v)
\nonumber \\
&= \big[x^* (z_1^{(\alpha)} \extk + w_1^{(\alpha)}) + y^*\big] + (u \extk + v) = (x^*z_1^{(\alpha)} \extk +  x^*w_1^{(\alpha)} + y^*) + (u \extk + v) 
\nonumber \\
&= (x^*z_1^{(\alpha)} + u)\extk +(  x^*w_1^{(\alpha)} + y^*+ v)
\end{align}
respectively.

\section{Algebraic properties of extended numbers } \label{extAlgebraSection}

In Abstract Algebra theory, the classification of any set of numbers plus the  sum and multiplication operations are described by the compliance or not of the following properties:
\begin{enumerate}
\item Associativity of addition and multiplication
\item Commutativity of addition and multiplication
\item Existence of additive and multiplicative identity elements
\item Existence of additive inverses and multiplicative inverses
\item Distributivity of multiplication over addition
\end{enumerate}
We start the study of the properties of the standard and conjugated sum and product of the extended numbers, using their definitions and taking into account that values like $z_0,w_0$ for the standard multiplication operation are parameters still to be determined.

\subsection*{The standard summation ``$+$''}
\begin{enumerate}
\item Associativity:
For all $\alpha_1 = x_1\extk + y_1, \alpha_2= x_2\extk + y_2, \alpha_3= x_3\extk + y_3 \; \in \mathbb{E}$ then
\begin{align}
&(\alpha_1 + \alpha_2) + \alpha_3  = \big[( x_1\extk + y_1) +  (x_2\extk + y_2)\big] +  (x_2\extk + y_3) 
\nonumber \\
&= \big[(x_1 + x_2)\extk + (y_1 + y_2)\big] +  (x_2\extk + y_3)
\nonumber \\
&= (x_1 + x_2 + x_3)\extk + (y_1 + y_2 + y_3) = x_1\extk + y_1 + \big[(x_2 + x_3)\extk + (y_2 + y_3)\big]
\nonumber \\
&= \alpha_1 + (\alpha_2 + \alpha_3)
\end{align}
\item Commutativity
For all $\alpha_1 = x_1\extk + y_1, \alpha_2= x_2\extk + y_2\; \in \mathbb{E}$ then
\begin{equation}
\alpha_1 + \alpha_2 = (x_1 + x_2)\extk + (y_1 + y_2) = \alpha_2 + \alpha_1 \qquad \forall \alpha_1,\alpha_2 \in \mathbb{E}
\end{equation}
\item Existence of additive identity element $0_E$ in $\mathbb{E}$ such that 
\begin{equation}
\alpha + 0_E = (x + 0)\extk + (y +0) = 0_E + \alpha = \alpha
\end{equation}
\item Existence of additive inverse such that for every $\alpha \in \; \mathbb{E}$, there exists an element $-\alpha\in\; \mathbb{E}$, such that $\alpha + (-\alpha) = 0$. If $\alpha = x \extk + y$ then $-\alpha = - x\extk -y$. It can be verified that
\begin{equation}
y \extk + y + (- x \extk -y) = (x - x)\extk + (y - y) = 0.
\end{equation}
\end{enumerate}

\subsection*{ The standard product ``$\cdot$''} 

\begin{enumerate}
\item Associativity:

Standard product is associative if the extended numbers $\alpha_1,\alpha_2,\alpha_3 \; \in \mathbb{E}$ satisfy
\begin{equation}
\alpha_1 \cdot (\alpha_2 \cdot \alpha_3) = (\alpha_1  \cdot \alpha_2) \cdot \alpha_3.
\end{equation}
Being $\alpha_1 = x_1\extk + y_1, \alpha_2= x_2\extk + y_2$ and $\alpha_3= x_3\extk + y_3 $. The left member of the above equation is
\begin{align}
& (x_1 \extk + y_1) \cdot \big[(x_2 \extk + y_2) \cdot (x_3 \extk + y_3)\big] = (x_1 \extk + y_1) \cdot \big[(x_2 x_3 z_0 + x_2 y_3 + y_2 x_3 )\extk + x_2 x_3 w_0 + y_2 y_3  \big]
\nonumber \\
&= (x_1 x_2 x_3 z_0 + x_1 x_2 y_3 + x_1 y_2 x_3 )\extk^2 + ( x_1 x_2 x_3 w_0 + x_1 y_2 y_3 + y_1 x_2 x_3 z_0 + y_1 x_2 y_3 + y_1 y_2 x_3 )\extk
\nonumber \\
&\quad + y_1 x_2 x_3 w_0 + y_1 y_2 y_3  
\nonumber \\
&= \big[x_1 x_2 x_3 z_0^2 + (x_1 x_2 y_3  + x_1 y_2 x_3 + y_1 x_2 x_3)z_0 + x_1 x_2 x_3 w_0 + x_1 y_2 y_3  + y_1 x_2 y_3 + y_1 y_2 x_3 \big]\extk 
\nonumber \\
& \quad + x_1 x_2 x_3 w_0 z_0  + (x_1 x_2 y_3  + x_1 y_2 x_3  + y_1 x_2 x_3) w_0 + y_1 y_2 y_3.
\end{align}
The right member of the axiomatic equation is
\begin{align}
& \big[(x_1 \extk + y_1) \cdot (x_2 \extk + y_2)\big]( \cdot x_3 \extk + y_3) = \big[ (x_1 x_2 z_0 + x_1 y_2 + y_1 x_2 )\extk + x_1 x_2 w_0 + y_1 y_2 \big] \cdot (x_3 \extk + y_3)
\nonumber \\
&= ( x_1 x_2 x_3 z_0 + x_1 y_2 x_3 + y_1 x_2 x_3 )\extk^2 + ( x_1 x_2 y_3 z_0 + x_1 y_2 y_3 + y_1 x_2 y_3  + x_1 x_2 x_3 w_0 + y_1 y_2 x_3)\extk
\nonumber \\
& \quad + x_1 x_2 y_3 w_0 + y_1 y_2 y_3 
\nonumber \\
&= \big[x_1 x_2 x_3 z_0^2 + (x_1 x_2 y_3  + x_1 y_2 x_3 + y_1 x_2 x_3)z_0 + x_1 x_2 x_3 w_0 + x_1 y_2 y_3  + y_1 x_2 y_3 + y_1 y_2 x_3 \big]\extk 
\nonumber \\
& \quad + x_1 x_2 x_3 w_0 z_0  + (x_1 x_2 y_3  + x_1 y_2 x_3  + y_1 x_2 x_3) w_0 + y_1 y_2 y_3.
\end{align}
which is the same result as above.
\item Distributivity of the standard multiplication over the standard addition

The standard extended inner product is distributive over the standard addition if for all $ \alpha_1,\alpha_2,\alpha_3 \in \mathbb{E}$ it is satisfied the relation 
\begin{equation}
\alpha_1 \cdot ( \alpha_2 + \alpha_3) = \alpha_1  \cdot \alpha_2 + \alpha_1  \cdot \alpha_3.
\end{equation}
If the numbers $\alpha_i = x_i \extk + y_i$, for all $i=1,2$, the right member of the preposition is
\begin{align}
&\alpha_1 \cdot ( \alpha_2 + \alpha_3)  = ( x_1 \extk + y_1 )  \cdot \big[ ( x_2 \extk + y_2 ) + ( x_3 \extk + y_3 )\big] = ( x_1 + \extk y_1 ) \big[ ( x_2 + x_3 )\extk  + y_2 + y_3 )\big]
\nonumber \\
&= \big[ x_1 ( x_2 + x_3 )z_0 + x_1 ( y_2 + y_3 ) + y_1 ( x_2 + x_3 ) \big]\extk + x_1 ( x_2 + x_3 )w_0 
\nonumber \\
& \quad + y_1 ( y_2 + y_3 )
\nonumber \\
&= \big[ ( x_1 x_2 z_0 + x_1 y_2 + y_1 x_2 )\extk + x_1 x_2 w_0 + y_1 y_2 \big] + \big[ ( x_1 x_3 z_0 + x_1 y_3 + y_1 x_3 )\extk 
\nonumber \\
& \quad + x_1 x_3 w_0 + y_1 y_3 \big]
\nonumber \\
& = ( x_1 \extk + y_1 ) \cdot ( x_2 \extk + y_2 ) + ( x_1 \extk + y_1 ) \cdot ( x_3 \extk + y_3 ) = \alpha_1 \cdot  \alpha_2 + \alpha_1  \cdot \alpha_3,
\end{align}
as stated before.
\item Commutativity:

The standard extended product is commutative if relation $\alpha_1 \alpha_2 = \alpha_2 \alpha_1$ is satisfied for all $ \alpha_1,\alpha_2 \in \mathbb{E}$. The replacement of numbers $\alpha_i = x_i \extk + y_i$, for all $i=1,2$, set standard product has the form:
\begin{align}
&\alpha_1  \cdot \alpha_2 = (x_1 \extk + y_1) \cdot (x_2 \extk + y_2) = (x_1 x_2 z_0 + x_1 y_2 + x_2 y_1)\extk + x_1 x_2 w_0 + y_1 y_2 
\nonumber \\
&=  (x_2 \extk + y_2) \cdot (x_1 \extk + y_1) = \alpha_2  \cdot \alpha_1
\end{align}
\item Zero-product property

In algebra, the zero-product property states that the product of two nonzero elements is nonzero. In other words, if:
\begin{equation}
\alpha  \cdot \beta = 0, \qquad \text{only if} \qquad \alpha = 0 \quad \text{or} \quad \beta =0.
\end{equation}
The standard product of these two extended numbers, where $\alpha=x\extk + y$ and $\beta =u \extk + v$, show the existence of nontrivial zero-divisors on the standard product. Setting zero the standard product of two extended numbers we have
\begin{equation}
\alpha  \cdot \beta = (x\extk + y) \cdot (u \extk + v)= \big[x(u z_0 + v) + y u\big]\extk + x u w_0 + y v=0
\end{equation}
which led to two complex equations
\begin{equation}
x(u z_0 + v) + y u =0, \qquad x u w_0 + y v=0 \label{zeroProEq}
\end{equation}
Without lost generality, we can analyze this set of equations for different cases of $\beta$ number:
\begin{enumerate}
\item $u=0, \;v\neq 0$.
In this case, the equations are
\begin{equation}
x v =0, \qquad y v=0
\end{equation}
which its satisfied if $x=y=0$.
\item $u\neq 0, \;v= 0$.
In this case, the equation \ref{zeroProEq} take the form
\begin{equation}
(x z_0 + y) u =0, \qquad x u w_0 = 0
\end{equation}
which its satisfied if $x=y=0$, for $w_0 \neq 0$. If $w_0 = 0$, then a nontrivial root from the equation $x z_0 + y=0$ is included.
\item $u\neq 0, \;v\neq 0$. In this case, multiplying first equation \ref{zeroProEq} by $v$, the second one by $u$ and subtracting one from the other we obtain
\begin{equation}
x(v^2 + u v z_0 - u_2 w_0)=0,
\end{equation}
which introduce the nontrivial root coming from the equation:
\begin{equation}
v^2 + u v z_0 - u_2 w_0=0
\end{equation}
\end{enumerate}
\item Existence of multiplicative identity element:

For every $\alpha \in \mathbb{E}$, exist a number   $1_E = 1$ in $\mathbb{E}$ such that 
\begin{equation}
\alpha \cdot  1 = 1  \cdot  \alpha = \alpha.
\end{equation}

\end{enumerate}

On the other side, the conjugate sum nor product is not associative, commutative, and has no additive identity nor inverse element. The following discussion is referred to the extended numbers whose maps $z_1$, $w_1$, $w_1$, and $w_2$ are defined. That means we exclude purely extended and complex numbers. For these cases, we should proceed using the extended unit definition.
\subsection*{The conjugated sum ``$\oplus$''}
\begin{enumerate}
\item Associativity

The associativity property implies that
\begin{equation}
(\alpha_1 \oplus \alpha_2)\oplus \alpha_3 =  \alpha_1 \oplus (\alpha_2 \oplus \alpha_3).
\end{equation}
The left member of the last relation
\begin{align}
&(\alpha_1 \oplus \alpha_2)\oplus \alpha_3 = \big[(x_1 \extk + y_1) \oplus (x_2 \extk + y_2)\big] \oplus (x_3 \extk + y_3)
\nonumber \\
&= \big[(x_1^* z_1^{(\alpha_1)} + x_2)\extk + x_1^* w_1^{(\alpha_1)} + y_1^* + y_2 \big]\oplus (x_3 \extk + y_3)
\nonumber \\
&= \big[ (x_1 z_1^{*(\alpha_1)} + x_2^*) z_1^{(\alpha_1\oplus \alpha_2)} + x_3 \big] \extk + (x_1 z_1^{*(\alpha_1)} + x_2^*) w_1^{(\alpha_1 \oplus \alpha_2)} + x_1^* w_1^{(\alpha_1)} + y_1^* + y_2
\end{align}
is not equal to the right member since
\begin{align}
& \alpha_1 \oplus (\alpha_2 \oplus \alpha_3) = (x_1 \extk + y_1) \oplus \big[(x_2 \extk + y_2) \oplus (x_3 \extk + y_3)\big] 
\nonumber \\
& = (x_1 \extk + y_1) \oplus \big[(x_2^* z_1^{(\alpha_2)} + x_3)\extk + x_2^* w_1^{(\alpha_2)} + y_2^* + y_3 \big]
\nonumber \\
&= \big[ x_1^* z_1^{(\alpha_1)} + x_2^* z_1^{(\alpha_2)} + x_3 \big] \extk + x_1^* w_1^{(\alpha_1)} +  x_2^* w_1^{(\alpha_2)} + y_1^* + y_2^* + y_3
\end{align}

\item Commutativity

The expression 
\begin{align}
&\alpha_1 \oplus \alpha_2 = (x_1 \extk + y_1) \oplus (x_2 \extk + y_2) 
\nonumber \\
&= \big[x_1^*z_1^{(\alpha_1)} + x_2\big]\extk + x_1^*w_1^{(\alpha_1)} + y_1^* + y_2
\end{align}
while
\begin{align}
&\alpha_2 \oplus \alpha_1 = (x_2 \extk + y_2) \oplus (x_1 \extk + y_1) 
\nonumber \\
&= \big[x_1 + x_2^*z_1^{(\alpha_2)}\big]\extk + x_2^*w_1^{(\alpha_2)} + y_1 + y_2^*,
\end{align}
which means that the conjugated sum does not satisfy the commutative property.
\item There is no conjugated additive identity element  $0_E$ in $\mathbb{E}$ because 
\begin{equation}
0_E \oplus \alpha =  x \extk + y  = \alpha
\end{equation}
while
\begin{equation}
\alpha \oplus 0_E = x^* z_1^{(\alpha)}\extk + x^* w_1^{(\alpha)} + y \neq \alpha
\end{equation}

\item The conjugated additive inverse element also does not exist. That can be probed by straight substitution and by noting that the property will not be satisfied if the conjugated sum is non-commutative.
\end{enumerate}

\subsection*{The conjugate product ``$\odot$''}
\begin{enumerate}
\item Associativity

The associativity property implies that
\begin{equation}
\alpha_1 \odot (\alpha_2 \odot \alpha_3) = (\alpha_1 \odot \alpha_2)\odot \alpha_3.
\end{equation}
Computing the left member we have
\begin{align}
&\alpha_1 \odot (\alpha_2\odot \alpha_3) = (x_1 \extk + y_1) \odot \big[(x_2 \extk + y_2) \odot (x_3 \extk + y_3)\big]
\nonumber \\
&= (x_1 \extk + y_1) \odot \big[(x_2^* y_3 z_1^{(\alpha_2)} + x_3 y_2^*)\extk + \imagi x_2^* x_3 + x_2^* y_3 w_1^{(\alpha_2)} + y_2^* y_3\big]
\nonumber \\
&=\big[x_1^*(\imagi x_2^* x_3 + x_2^* y_3 w_1^{(\alpha_2)} + y_2^* y_3) z_1^{(\alpha_1)} + y_1^*(x_2^* y_3 z_1^{(\alpha_2)} + x_3 y_2^*) \big] \extk 
\nonumber \\
& \quad + \imagi x_1^*(x_2^* y_3 z_1^{(\alpha_2)} + x_3 y_2^*) + x_1^*(\imagi x_2^* x_3 + x_2^* y_3 w_1^{(\alpha_2)} + y_2^* y_3) w_1^{(\alpha_1)} 
\nonumber \\ 
& \quad + y_1^*(\imagi x_2^* x_3 + x_2^* y_3 w_1^{(\alpha_2)} + y_2^* y_3)
\nonumber \\ 
&=\big[\imagi x_1^* x_2^* x_3 z_1^{(\alpha_1)}+ x_1^* x_2^* y_3 z_1^{(\alpha_1)} w_1^{(\alpha_2)} + x_1^* y_2^* y_3z_1^{(\alpha_1)}  + x_2^* y_1^* y_3 z_1^{(\alpha_2)} + x_3 y_1^*y_2^* \big] \extk 
\nonumber \\
& \quad + x_1^*x_2^* y_3 z_1^{(\alpha_2)} + x_1^* x_3 y_2^* + x_1^*\imagi x_2^* x_3w_1^{(\alpha_1)} + x_1^*x_2^* y_3 w_1^{(\alpha_1)} w_1^{(\alpha_2)} + x_1^*y_2^* y_3 w_1^{(\alpha_1)} 
\nonumber \\ 
& \quad + \imagi x_2^* x_3 y_1^*+ x_2^* y_1^* y_3 w_1^{(\alpha_2)} + y_1^*y_2^* y_3.
\end{align}
while the computation of the right member is 
\begin{align}
&(\alpha_1 \odot \alpha_2)\odot \alpha_3 = \big[(x_1 \extk + y_1) \odot (x_2 \extk + y_2)\big] \odot (x_3 \extk + y_3)
\nonumber \\
&=\big[(x_1^* y_2 z_1^{(\alpha_1)} + x_2 y_1^*)\extk  + \imagi x_1^* x_2 + x_1^* y_2 w_1^{(\alpha_1)} + y_1^* y_2\big]\odot (x_3 \extk + y_3)
\nonumber \\
&= \big[ (x_1^* y_2 z_1^{(\alpha_1)} + x_2 y_1^*)^*y_3z_1^{(\alpha_1 \odot \alpha_2)} + (\imagi x_1^* x_2 + x_1^* y_2 w_1^{(\alpha_1)} + y_1^* y_2)^*x_3\big]\extk
\nonumber \\
&\quad + \imagi (x_1^* y_2 z_1^{(\alpha_1)} + x_2 y_1^*)^* x_3  + (x_1^* y_2 z_1^{(\alpha_1)} + x_2 y_1^*)^*y_3w_1^{(\alpha_1 \odot \alpha_2)}  
\nonumber \\
&\quad +  (\imagi x_1^* x_2 + x_1^* y_2 w_1^{(\alpha_1)} + y_1^* y_2)^*y_3
\nonumber \\
&= \big[ x_1 y_2^*y_3 z_1^{*(\alpha_1)}z_1^{(\alpha_1 \odot \alpha_2)} 
+ x_2^* y_1 y_3z_1^{(\alpha_1 \odot \alpha_2)} 
- \imagi x_1 x_2^*x_3 + x_1 x_3 y_2^* w_1^{*(\alpha_1)} + x_3y_1 y_2^*\big]\extk
\nonumber \\
&\quad + \imagi x_1 x_3 y_2^* z_1^{*(\alpha_1)} + \imagi x_2^* x_3 y_1 + x_1 y_2^*y_3 z_1^{*(\alpha_1)}w_1^{(\alpha_1 \odot \alpha_2)} 
+ x_2^* y_1 y_3 w_1^{(\alpha_1 \odot \alpha_2)} 
\nonumber \\
&\quad   \imagi x_1 x_2^*y_3 + x_1 y_2^*y_3 w_1^{*(\alpha_1)} + y_1 y_2^*y_3,
\end{align}
showing that
\begin{equation}
\alpha_1 \odot (\alpha_2 \odot \alpha_3) \neq (\alpha_1 \odot \alpha_2)\odot \alpha_3,
\end{equation}
or what is the same, it not comply with the associative property.
\item Commutativity

The commutative property, $\alpha_1 \odot \alpha_2 = \alpha_2 \odot \alpha_1$, $ \forall \alpha_1,\alpha_2 \in \mathbb{E}$ is not satisfied. For numbers  $\alpha_i = x_i \extk + y_i$, where $i=1,2$, we have:
\begin{align}
\alpha_1 \odot \alpha_2 = (x_1^* y_2 z_1^{(\alpha_1)} + x_2 y_1^*)\extk + x_1^* y_2 w_1^{(\alpha_1)} + \imagi x_1^* x_2 + y_1^* y_2 
\end{align}
while 
\begin{align}
\alpha_2 \odot \alpha_1 = (x_2^* y_1 z_1^{(\alpha_2)} + x_1 y_2^*)\extk + x_2^* y_1 w_1^{(\alpha_2)} + \imagi x_2^* x_1 + y_1 y_2^*
\end{align}

Due to the noncommutative property of the conjugated product, it's convenient to specify the order of the conjugated multiplication. Then, we can assume that the left conjugate multiplication of an extended number  $\alpha$ by other extended $\beta$ stands for  $\alpha  \odot \beta$ the right multiplication of  $\alpha$ by $\beta$ means $ \beta \odot \alpha$.

\item Distributivity of the conjugated multiplication over the standard addition

We study the distributive property for the right and left conjugated multiplication of the standard sum. In the first case, this property is satisfied if:
\begin{equation}
\alpha_1 \odot ( \alpha_2 + \alpha_3) = \alpha_1 \odot \alpha_2 + \alpha_1 \odot \alpha_3.
\end{equation}
Computing the left member we have
\begin{align}
&\alpha_1 \odot ( \alpha_2 + \alpha_3)  = (x_1 \extk + y_1) \odot \big[(x_2 \extk + y_2) + (x_3 \extk + y_3)\big]
\nonumber \\
&= \big[ x_1^*(y_2 + y_3) z_1^{(\alpha_1)} + (x_2 + x_3)y_1^* \big] \extk + x_1^*(y_2 + y_3) w_1^{(\alpha_1)} + \imagi  x_1^*(x_2 + x_3) +  y_1^*(y_2 + y_3)
\nonumber \\
&=  \big[ x_1^*y_2  z_1^{(\alpha_1)} + x_2 y_1^* \big] \extk + x_1^* y_2 w_1^{(\alpha_1)} + \imagi  x_1^*x_2 +  y_1^* y_2
\nonumber \\
&\quad + \big[ x_1^*y_3  z_1^{(\alpha_1)} + x_3 y_1^* \big] \extk + x_1^* y_3 w_1^{(\alpha_1)} + \imagi  x_1^*x_3 +  y_1^* y_3
\nonumber \\
&=  \alpha_1 \odot \alpha_2 + \alpha_1 \odot \alpha_3,
\end{align}
showing that the right conjugated product is distributive. Instead, we cannot arrive at the same conclusion for the left conjugated multiplication of the sum. The distributivity property for the left conjugated multiplication of the sum is verified if:
\begin{equation}
(\alpha_1 + \alpha_2) \odot \alpha_3 = \alpha_1 \odot \alpha_3 + \alpha_2 \odot \alpha_3.
\end{equation}
The left member of the previous expression
\begin{align}
&(\alpha_1 + \alpha_2) \odot \alpha_3 = \big[(x_1 \extk + y_1) + (x_2 \extk + y_2)\big] \odot (x_3 \extk + y_3)
\nonumber \\
&=\big[(x_1  + x_2 )\extk + (y_1 + y_2)\big] \odot (x_3 \extk + y_3)\big]
\nonumber \\
&= \big[(x_1^*  + x_2^* )y_3 z_1^{(\alpha_1 + \alpha_2)}  + x_3(y_1^* + y_2^*)\big]\extk + (x_1^*  + x_2^* )y_3 w_1^{(\alpha_1 + \alpha_2)} 
\nonumber \\
&\quad + \imagi (x_1^*  + x_2^* )x_3 +  (y_1^* + y_2^*)y_3
\end{align}
is different from the right's
\begin{align}
&\alpha_1\odot \alpha_3 + \alpha_2 \odot \alpha_3 = (x_1 \extk + y_1) \odot (x_3 \extk + y_3) + (x_2 \extk + y_2) \odot (x_3 \extk + y_3)
\nonumber \\
&=\big[(x_1^*z_1^{(\alpha_1)}  + x_2^*z_1^{(\alpha_2)} )y_3 + x_3(y_1^* + y_2^*) \big]\extk +  (x_1^*w_1^{(\alpha_1)}  + x_2^*w_1^{(\alpha_2)} )y_3 
\nonumber \\
& \quad + \imagi (x_1^* + x_2^*)x_3 + (y_1^* + y_2^*)y_3.
\end{align}
The distributive property is then satisfied if :
\begin{align}
\mathcal{D}^{(\alpha_1 + \alpha_2)}(\alpha_1,\alpha_2,\alpha_3) &\equiv \big[ (x_1^*  + x_2^* )y_3 z_1^{(\alpha_1 + \alpha_2)} - (x_1^*z_1^{(\alpha_1)}  + x_2^*z_1^{(\alpha_2)} )y_3 \big ] \extk + 
\nonumber \\
& \qquad (x_1^*  + x_2^* )y_3 w_1^{(\alpha_1 + \alpha_2)} - (x_1^*w_1^{(\alpha_1)}  + x_2^*w_1^{(\alpha_2)} )y_3 = 0  \label{extDfuntion}
\end{align}
is zero. The function $\mathcal{D}^{(\alpha_1 + \alpha_2)}(\alpha_1,\alpha_2,\alpha_3)$ measure the magnitude of the differences between the factors of the maps. Subtracting the explicit product of both members, the distribution law can be expressed as:
\begin{equation}
(\alpha_1 + \alpha_2) \odot \alpha_3 = \alpha_1 \odot \alpha_3 + \alpha_2 \odot \alpha_3 + 
\mathcal{D}^{(\alpha_1 + \alpha_2)}(\alpha_1,\alpha_2,\alpha_3). \label{distPropRight}
\end{equation}

The function $\mathcal{D}^{(\alpha_1 + \alpha_2)}(\alpha_1,\alpha_2,\alpha_3)$ is null if
\begin{itemize}
\item the maps  for $\alpha_1$ and $\alpha_2$ numbers satisfy:
\begin{equation}
z_1^{(\alpha_1)} = z_1^{(\alpha_2)} =  z_1^{(\alpha_1 + \alpha_2)}, \qquad
w_1^{(\alpha_1)} = w_1^{(\alpha_2)} =  w_1^{(\alpha_1 + \alpha_2)} 
\end{equation}
\item $\alpha_1$ and $\alpha_2$  are both pure complex numbers, $e.i.$ $x_1 = x_2 = 0$
\item  $\alpha_3$ is a pure complex numbers,  $e.i.$ $y_3 =0$
\end{itemize}
It can also be verified that
\begin{equation}
\mathcal{D}^{(\alpha_1 + \alpha_2)}(\alpha_1,\alpha_2,\alpha_3) 
+ \mathcal{D}^{(\alpha_1 + \alpha_2)}(\alpha_1,\alpha_2,\alpha_4) 
= \mathcal{D}^{(\alpha_1 + \alpha_2)}(\alpha_1,\alpha_2,\alpha_3 + \alpha_4).\label{distPropRight1}
\end{equation}

\item Existence of the multiplicative identity element.

If the conjugate product has an identity element $1_E^{(\oplus)}$, it must satisfy
\begin{equation}
\alpha \odot 1_E^{(\oplus)} = 1_E^{(\oplus)} \odot \alpha = \alpha.
\end{equation}
From the noncommutative property, we can see that the identity element does not exist. 
\end{enumerate}

We do not attempt in here to classify the set of the extended numbers; however, according to Abstract Algebra, they behave like a commutative ring with the existence of nontrivial zero divisors. We note that, same as extended numbers that satisfy the Associativity, Commutativity, and Distributivity axioms for the standard sum and product, the pair $\alpha \odot \beta$ will also satisfy the same properties. That means, for example, that pairs satisfy the distribution property:
\begin{equation}
(\alpha_1 \odot \beta_1) \cdot \big[(\alpha_2 \odot \beta_2) + (\alpha_3 \odot \beta_3)\big] = (\alpha_1 \odot \beta_1) \cdot (\alpha_2 \odot \beta_2) + (\alpha_1 \odot \beta_1) \cdot (\alpha_3 \odot \beta_3).\label{ExtVectorDistrProp}
\end{equation}

Also, a product like 
\begin{equation}
(\alpha_1^\bullet \odot \beta_1^\bullet)\cdot (\alpha_2 \odot \beta_2)
\end{equation}
satisfies the distributive properties
\begin{equation}
(\alpha_1^\bullet \odot \beta_1^\bullet)\cdot \big[(\alpha_2 \odot \beta_2) + (\alpha_3 \odot \beta_3)\big] 
= (\alpha_1^\bullet \odot \beta_1^\bullet) \cdot (\alpha_2 \odot \beta_2) + (\alpha_1^\bullet \odot \beta_1^\bullet) \cdot (\alpha_3 \odot \beta_3) \label{ExtProdDistrProp1}
\end{equation}
and
\begin{equation}
\big[(\alpha_1^\bullet \odot \beta_1^\bullet) + (\alpha_2^\bullet \odot \beta_2^\bullet) \big] \cdot  (\alpha_3 \odot \beta_3) = (\alpha_1^\bullet \odot \beta_1^\bullet) \cdot (\alpha_3 \odot \beta_3) + (\alpha_2^\bullet \odot \beta_2^\bullet) \cdot (\alpha_3 \odot \beta_3).\label{ExtProdDistrProp2}
\end{equation}
The properties for quantities $(\alpha \odot \beta)$, together with the intuition of the form of a quantum operator, act over a two-component quantum state for $n$-VMVF systems, give us a hint for finding the form of the inner product in the extended domain.

\section{The extended inner product.} \label{inner_prod_chapter}
We are now able to propose the absolute value for an extended number using the new operations.

The inner product should be a set of operations applied on four extended numbers, according to the first section of this chapter, and also connected with the two types of products. The most straightforward possible definitions for the inner product for the extended numbers $\alpha , \beta , \gamma , \delta$ are, regardless of the order of priority,
\begin{align}
&1.\quad \alpha^\bullet \cdot \beta^\bullet \cdot \gamma \cdot \delta   &&5.\quad \alpha^\bullet \odot \beta^\bullet \cdot \gamma \cdot \delta 
\nonumber \\
&2.\quad \alpha^\bullet \cdot \beta^\bullet \cdot \gamma \odot \delta   &&6.\quad \alpha^\bullet \odot \beta^\bullet \cdot \gamma \odot \delta 
\nonumber \\
&3.\quad \alpha^\bullet \cdot \beta^\bullet \odot \gamma \cdot \delta   &&7.\quad \alpha^\bullet \odot \beta^\bullet \odot \gamma \cdot \delta 
\nonumber \\
&4.\quad \alpha^\bullet \cdot \beta^\bullet \odot \gamma \odot \delta   &&8.\quad \alpha^\bullet \odot \beta^\bullet \odot \gamma \odot \delta, 
\end{align}
where we include the extended conjugate map $()^\bullet$.

As mentioned in the sections ``Introduction'', the primary motivation of this work is the construction of a Hilbert space over a new domain of numbers that includes negative probabilities and, most important, that fits the new structure obtained in the classical theory. The Hamilton theory shows that two canonical transformations are needed to evolve the system, one using the rectangular coordinates and another using the angular's \cite{Israel:1811.12175}. A point in the canonical space and with it, a state of the system has two components: the rectangular and the angular. The extended \textit{ket} should have then two components in correspondence with the canonical space, which, together with an extended \textit{bra}, that should also have two components, define the inner product for obtaining real measurements. That points out that there should exist some pair-pair symmetry in the inner product. The cases 1,3,6 and 8 are the only ones who have that kind of symmetry. The superposition principle also points out that the state vectors must satisfy the Associativity, Commutativity, and Distributivity properties. In that case, the product operation that satisfies those properties is the standard multiplication, retaining only cases 1 and 6. The first proposition for the inner product, do not include the complex product or the new map $()^\bullet$.
 Because of that, according to the discussed above, it will not result in a real number.

Based on this weak explanation embedded with intuition and the need to adjust the inner product to the expected behavior for the quantum theory, we propose the definition of the inner product of four extended numbers $\alpha , \beta , \gamma , \delta$ as:
\begin{equation}
\langle \alpha , \beta , \gamma , \delta \rangle \equiv (\alpha^\bullet \odot \beta^\bullet) \cdot (\gamma \odot \delta). \label{ExtInnerProdDef}
\end{equation}
The four power of the absolute value for an extended number $\alpha = x \extk + y$ can be then written as:
\begin{align}
|\alpha|^4 &= ( \alpha^\bullet \odot \alpha^\bullet )\cdot (\alpha \odot \alpha )=
\\
&=\big[(x \extk + y)^{\bullet *} (x \extk + y)^\bullet\big] \cdot \big[(x \extk + y)^* (x \extk + y)\big], \;\;\; \forall x,y \in \mathbb{C}.
\end{align}
for  $x=1$ and $y=0$ we rewrite the first equation for the absolute value of the extended unit, which we supposed is equal to $1$:
\begin{equation}
\big(\extk^\bullet \odot \extk^\bullet\big)\cdot \big(\extk \odot \extk\big)=1
\end{equation}

We proposed that the extended unit definition stands for all the extended numbers, \textit{e.i.} $\extk \odot \extk= \imagi$. This proposition, with the absolute value of the extended unit equal to 1, set a constraint:
\begin{equation}
\extk^\bullet \odot \extk^\bullet = \imagi^*.
\end{equation}
We can consider this equation to stand with no loss of generality for all the extended numbers
when we have a product like:
\begin{equation}
\alpha^\bullet \odot \beta^\bullet = (\alpha_E \extk  + \alpha_I)^\bullet \odot (\beta_E \extk  + \beta_I)^\bullet 
= (\alpha_E \extk^{(\alpha)^\bullet}  + \alpha_I) \odot (\beta_E \extk^{(\beta)^\bullet}  + \beta_I) 
\end{equation}

In the case of $\alpha^\bullet \odot \alpha^\bullet$, we have:
\begin{align}
\extk^{(\alpha)^\bullet} \odot \extk^{(\alpha)^\bullet} &= (z_2^{(\alpha)} \extk + w_2^{(\alpha)}) \odot (z_2^{(\alpha)} \extk + w_2^{(\alpha)})
\nonumber \\
&= (z_1^{(\alpha^\bullet)} {z_2^{(\alpha)}}^* w_2^{(\alpha)} + z_2^{(\alpha)} {w_2^{(\alpha)}}^*)\extk + \imagi |z_2^{(\alpha)}|^2 + {z_2^{(\alpha)}}^* w_1^{(\alpha^\bullet)} w_2^{(\alpha)} + |w_2^{(\alpha)}|^2 = \imagi^* \label{extConjDangerous}
\end{align}
which lead to the relations between the coefficients
\begin{align}
z_1^{(\alpha^\bullet)} {z_2^{(\alpha)}}^* w_2^{(\alpha)} + z_2^{(\alpha)} {w_2^{(\alpha)}}^* &= 0
\nonumber \\
-|z_2^{(\alpha)}|^2 
+ \imagi {z_2^{(\alpha)}}^* w_1^{(\alpha^\bullet)} w_2^{(\alpha)} 
+ \imagi|w_2^{(\alpha)}|^2 &= 1 \label{extConjRelations}.
\end{align}
Note the appearance of coefficients related to the number $\alpha^\bullet$.


Following the rules of the standard and conjugated product, we have the expression of the absolute value power four:
\begin{equation}
|\alpha|^4 =\Gamma_E \extk + \Gamma_I, \label{extFinalAbsRelations}
\end{equation}
where
\begin{align}
&\Gamma_E = 
|x|^4 \Big(
\imagi z_1^{(\alpha^\bullet)} {z_2^{(\alpha)}}^* w_2^{(\alpha)} 
+ \imagi z_2^{(\alpha)} {w_2^{(\alpha)}}^*
\Big)
\nonumber \\
&+|x|^2|y|^2 \Big(
  z_0 z_1^{(\alpha)} z_2^{(\alpha)} 
+ z_0 z_1^{(\alpha^\bullet)} {z_2^{(\alpha)} }^*
+ z_1^{(\alpha^\bullet)}{z_2^{(\alpha)}}^* w_2^{(\alpha)}
+ z_2^{(\alpha)} {w_2^{(\alpha)} }^*
+ z_2^{(\alpha)} w_1^{(\alpha)}
\nonumber \\
&\quad + z_1^{(\alpha)} w_2^{(\alpha)}
+ {w_2^{(\alpha)}}^*
+ {z_2^{(\alpha)}}^* w_1^{(\alpha^\bullet)} 
\Big) 
\nonumber \\
&+ |x|^2 x^*y \Big(
z_0 z_1^{(\alpha^\bullet)} {z_1^{(\alpha)} }^* z_2^{(\alpha)} w_2^{(\alpha)}
+ z_0 z_1^{(\alpha)} z_2^{(\alpha)} {w_2^{(\alpha)}}^*
+ \imagi z_1^{(\alpha^\bullet)} {z_2^{(\alpha)}}^* 
+ z_1^{(\alpha^\bullet)} {z_2^{(\alpha)}}^* w_1^{(\alpha)} w_2^{(\alpha)}
\nonumber \\
&\quad + z_2^{(\alpha)} w_1^{(\alpha)} {w_2^{(\alpha)}}^*
+ \imagi z_1^{(\alpha)} |z_2^{(\alpha)}|^2
+ z_1^{(\alpha)} |w_2^{(\alpha)}|^2
+ z_1^{(\alpha)} {z_2^{(\alpha)}}^* w_1^{(\alpha^\bullet)}
\Big)
\nonumber \\
&+ |x|^2 xy^* \Big(
z_0 z_1^{(\alpha^\bullet)} {z_2^{(\alpha)}}^* w_2^{(\alpha)}
+ z_0 z_2^{(\alpha)} {w_2^{(\alpha)} }^*
+ \imagi z_2^{(\alpha)}
+ \imagi |z_2^{(\alpha)}|^2
+ |w_2^{(\alpha)}|^2
+ {z_2^{(\alpha)}}^* w_1^{(\alpha^\bullet)} w_2^{(\alpha)}
\Big) 
\nonumber \\
&+ |y|^2 x^*y \Big(
z_1^{(\alpha^\bullet)} {z_2^{(\alpha)}}^* 
+ z_1^{(\alpha)}
\Big) 
\nonumber \\
&+ |y|^2 xy^*\Big(
z_2^{(\alpha)} + 1
\Big) 
\nonumber \\
&+ (xy^*)^2\Big(
z_0 z_2^{(\alpha)} 
+ w_2^{(\alpha)}
\Big)
\nonumber \\
&+ (x^*y)^2\Big(
z_0 z_1^{(\alpha^\bullet)} z_1^{(\alpha)} {z_2^{(\alpha)}}^*
+ z_1^{*(\alpha^\bullet)} {z_2^{(\alpha)}}^* w_1^{(\alpha)} 
+ z_1^{*(\alpha)} {z_2^{(\alpha)}}^* w_1^{(\alpha^\bullet)} 
+ z_1^{*(\alpha)} {w_2^{(\alpha)}}^*
\Big)
\label{extFinalAbsRelations_E}
\end{align}
and
\begin{align}
&\Gamma_I = 
|x|^4 
\Big( 
-|z_2^{(\alpha)}|^2 
+ \imagi {z_2^{(\alpha)}}^* w_1^{(\alpha^\bullet)} w_2^{(\alpha)} 
+ \imagi|w_2^{(\alpha)}|^2
\Big)
+ |y|^4 
\nonumber \\
&+ |x|^2|y|^2
\Big(
  z_1^{(\alpha)} z_2^{(\alpha)}  w_0 
+ z_1^{(\alpha^\bullet)} {z_2^{(\alpha)}}^* w_0 
+ \imagi
+ \imagi |z_2^{(\alpha)}|^2 
+ |w_2^{(\alpha)}|^2 
+ {z_2^{(\alpha)}}^* w_1^{(\alpha^\bullet)} w_2^{(\alpha)}
+ w_1^{(\alpha)} w_2^{(\alpha)} 
\Big) 
\nonumber \\
&+ |x|^2 x^*y
\Big(
z_1^{(\alpha^\bullet)} {z_1^{(\alpha)} }^* z_2^{(\alpha)} w_2^{(\alpha)} w_0
+ z_1^{(\alpha)} z_2^{(\alpha)} {w_2^{(\alpha)}}^* w_0
+ \imagi {w_2^{(\alpha)}}^* 
+ \imagi {z_2^{(\alpha)}}^* w_1^{*(\alpha^\bullet)} 
+ \imagi |z_2^{(\alpha)}|^2 w_1^{*(\alpha)} 
\nonumber \\
& \quad + w_1^{(\alpha)} |w_2^{(\alpha)} |^2
+ {z_2^{(\alpha)}}^* w_1^{(\alpha)}  w_1^{(\alpha^\bullet)} w_2^{(\alpha)} 
\Big) 
\nonumber \\
&+  |x|^2 xy^*
\Big(
z_1^{(\alpha^\bullet)} {z_2^{(\alpha)}}^* w_2^{(\alpha)} w_0 
+ z_2^{(\alpha)} {w_2^{(\alpha)} }^* w_0
+ \imagi w_2^{(\alpha)}
\Big) 
\nonumber \\
&+ |y|^2 x^*y
\Big(
{z_2^{(\alpha)}}^* w_1^{(\alpha^\bullet)} 
+ {w_2^{(\alpha)}}^* 
+ w_1^{(\alpha)}
\Big)
\nonumber \\
&+ |y|^2 xy^*
\Big(
w_2^{(\alpha)}
\Big)
\nonumber \\
&+ (x^*y)^2
\Big(
z_1^{(\alpha^\bullet)} z_1^{(\alpha)} {z_2^{(\alpha)}}^* w_0 
+ {z_2^{(\alpha)}}^*  w_1^{(\alpha^\bullet)} w_1^{(\alpha)}
+ {w_2^{(\alpha)}}^*
\Big) 
\nonumber \\
&+ (xy^*)^2
\Big(z_2^{(\alpha)} w_0
\Big).
\label{extFinalAbsRelations_I}
\end{align}
We can explicitly replace the first line of expression for the extended part and the imaginary part of the absolute value power four, using the equations \ref{extConjRelations}, like:
\begin{align}
|x|^4 \Big(
\imagi z_1^{(\alpha^\bullet)} {z_2^{(\alpha)}}^* w_2^{(\alpha)} 
+ \imagi z_2^{(\alpha)} {w_2^{(\alpha)}}^*
\Big) &= 0
\nonumber \\
|x|^4 
\Big( 
-|z_2^{(\alpha)}|^2 
+ \imagi {z_2^{(\alpha)}}^* w_1^{(\alpha^\bullet)} w_2^{(\alpha)} 
+ \imagi|w_2^{(\alpha)}|^2
\Big)
+ |y|^4  &= |x|^4  + |y|^4, 
\end{align}
respectively. We can also group the terms of the result using the parameters  $\phi = \frac{|x|}{|y|}$ and $\theta = \theta_x - \theta_y$, where  $|x|, \theta_x, |y| $ and $ \theta_y$ are the absolute values and angles of complex numbers $x$ and $y$, respectively. The following expressions can be written as:
\begin{align}
(x^*y)^2   &= |x|^2 |y|^2 e^{- 2\imagi \theta}
\nonumber \\
(xy^*)^2   &= |x|^2 |y|^2 e^{2\imagi \theta}
\nonumber \\
|x|^2 x^*y &= |x|^2 |y|^2 \phi e^{-\imagi \theta}
\nonumber \\
|x|^2 xy^* &= |x|^2 |y|^2 \phi e^{ \imagi \theta}
\nonumber \\
|y|^2 x^*y &= |x|^2 |y|^2 \phi^- e^{-\imagi \theta}
\nonumber \\
|y|^2 xy^* &= |x|^2 |y|^2 \phi^- e^{\imagi \theta}. \label{extFinalParam}
\end{align}
Replacing them, we obtain
\begin{align}
&\Gamma_E = 
|x|^2|y|^2 \Big[ 
  z_0 z_1^{(\alpha)} z_2^{(\alpha)} 
+ z_0 z_1^{(\alpha^\bullet)} {z_2^{(\alpha)} }^*
+ z_1^{(\alpha^\bullet)}{z_2^{(\alpha)}}^* w_2^{(\alpha)}
+ z_2^{(\alpha)} {w_2^{(\alpha)} }^*
+ z_2^{(\alpha)} w_1^{(\alpha)}
+ z_1^{(\alpha)} w_2^{(\alpha)}
\nonumber \\
&+ {w_2^{(\alpha)}}^*
+ {z_2^{(\alpha)}}^* w_1^{(\alpha^\bullet)} 
+ \phi e^{-\imagi \theta} \Big(
z_0 z_1^{(\alpha^\bullet)} {z_1^{(\alpha)} }^* z_2^{(\alpha)} w_2^{(\alpha)}
+ z_0 z_1^{(\alpha)} z_2^{(\alpha)} {w_2^{(\alpha)}}^*
+ \imagi z_1^{(\alpha^\bullet)} {z_2^{(\alpha)}}^* 
\nonumber \\
&+ z_1^{(\alpha^\bullet)} {z_2^{(\alpha)}}^* w_1^{(\alpha)} w_2^{(\alpha)}
+ z_2^{(\alpha)} w_1^{(\alpha)} {w_2^{(\alpha)}}^*
+ \imagi z_1^{(\alpha)} |z_2^{(\alpha)}|^2
+ z_1^{(\alpha)} |w_2^{(\alpha)}|^2
+ z_1^{(\alpha)} {z_2^{(\alpha)}}^* w_1^{(\alpha^\bullet)}
\Big)
\nonumber \\
&+ \phi e^{ \imagi \theta} \Big(
z_0 z_1^{(\alpha^\bullet)} {z_2^{(\alpha)}}^* w_2^{(\alpha)}
+ z_0 z_2^{(\alpha)} {w_2^{(\alpha)} }^*
+ \imagi z_2^{(\alpha)}
+ \imagi |z_2^{(\alpha)}|^2
+ |w_2^{(\alpha)}|^2
+ {z_2^{(\alpha)}}^* w_1^{(\alpha^\bullet)} w_2^{(\alpha)}
\Big) 
\nonumber \\
&+ \phi^- e^{-\imagi \theta} \Big(
z_1^{(\alpha^\bullet)} {z_2^{(\alpha)}}^* 
+ z_1^{(\alpha)}
\Big) 
+ \phi^- e^{\imagi \theta}\Big(
z_2^{(\alpha)} + 1
\Big) 
+ e^{2\imagi \theta}\Big(
z_0 z_2^{(\alpha)} 
+ w_2^{(\alpha)}
\Big)
\nonumber \\
&+ e^{- 2\imagi \theta}\Big(
z_0 z_1^{(\alpha^\bullet)} z_1^{(\alpha)} {z_2^{(\alpha)}}^*
+ z_1^{*(\alpha^\bullet)} {z_2^{(\alpha)}}^* w_1^{(\alpha)} 
+ z_1^{*(\alpha)} {z_2^{(\alpha)}}^* w_1^{(\alpha^\bullet)} 
+ z_1^{*(\alpha)} {w_2^{(\alpha)}}^*
\Big)
\Big]
\label{extFinalAbsRelations_E_1}
\end{align}
and
\begin{align}
&\Gamma_I = 
|x|^4 + |y|^4  + |x|^2|y|^2
\Big[
  z_1^{(\alpha)} z_2^{(\alpha)}  w_0 
+ z_1^{(\alpha^\bullet)} {z_2^{(\alpha)}}^* w_0 
+ \imagi
+ \imagi |z_2^{(\alpha)}|^2 
+ |w_2^{(\alpha)}|^2 
+ {z_2^{(\alpha)}}^* w_1^{(\alpha^\bullet)} w_2^{(\alpha)}
\nonumber \\
&+ w_1^{(\alpha)} w_2^{(\alpha)} 
+ \phi e^{-\imagi \theta}
\Big(
z_1^{(\alpha^\bullet)} {z_1^{(\alpha)} }^* z_2^{(\alpha)} w_2^{(\alpha)} w_0
+ z_1^{(\alpha)} z_2^{(\alpha)} {w_2^{(\alpha)}}^* w_0
+ \imagi {w_2^{(\alpha)}}^* 
+ \imagi {z_2^{(\alpha)}}^* w_1^{*(\alpha^\bullet)} 
\nonumber \\
&+ \imagi |z_2^{(\alpha)}|^2 w_1^{*(\alpha)} 
+ w_1^{(\alpha)} |w_2^{(\alpha)} |^2
+ {z_2^{(\alpha)}}^* w_1^{(\alpha)}  w_1^{(\alpha^\bullet)} w_2^{(\alpha)} 
\Big) 
+  \phi e^{ \imagi \theta}
\Big(
z_1^{(\alpha^\bullet)} {z_2^{(\alpha)}}^* w_2^{(\alpha)} w_0 
+ z_2^{(\alpha)} {w_2^{(\alpha)} }^* w_0
\nonumber \\
&+ \imagi w_2^{(\alpha)}
\Big) 
+ \phi^- e^{-\imagi \theta}
\Big(
{z_2^{(\alpha)}}^* w_1^{(\alpha^\bullet)} 
+ {w_2^{(\alpha)}}^* 
+ w_1^{(\alpha)}
\Big)
+ \phi^- e^{\imagi \theta}
\Big(
w_2^{(\alpha)}
\Big)
+ e^{- 2\imagi \theta}
\Big(
z_1^{(\alpha^\bullet)} z_1^{(\alpha)} {z_2^{(\alpha)}}^* w_0 
\nonumber \\
&+ {z_2^{(\alpha)}}^*  w_1^{(\alpha^\bullet)} w_1^{(\alpha)}
+ {w_2^{(\alpha)}}^*
\Big) 
+ e^{2\imagi \theta}
\Big(z_2^{(\alpha)} w_0
\Big)
\Big].
\label{extFinalAbsRelations_I_1}
\end{align}

The positive-definiteness condition of the absolute value state that the extended part of the extended equation \ref{extFinalAbsRelations} must be zero, and the imaginary part must be real and greater than zero. Applying that conjecture to our results and putting, together with relations \ref{extConjRelations}, we obtain the set of equations that relate the coefficients $z_1^{(\alpha)},w_1^{(\alpha)}$ and $z_2^{(\alpha)}, w_2^{(\alpha)}$:

\begin{align}
&\mathbf{1}. \; z_1^{(\alpha^\bullet)} {z_2^{(\alpha)}}^* w_2^{(\alpha)} + z_2^{(\alpha)} {w_2^{(\alpha)}}^* = 0
\nonumber \\
&\mathbf{2}. \; -|z_2^{(\alpha)}|^2 
+ \imagi {z_2^{(\alpha)}}^* w_1^{(\alpha^\bullet)} w_2^{(\alpha)} 
+ \imagi|w_2^{(\alpha)}|^2 = 1 
\nonumber \\
&\mathbf{3}. \;   z_0 z_1^{(\alpha)} z_2^{(\alpha)} 
+ z_0 z_1^{(\alpha^\bullet)} {z_2^{(\alpha)} }^*
+ z_1^{(\alpha^\bullet)}{z_2^{(\alpha)}}^* w_2^{(\alpha)}
+ z_2^{(\alpha)} {w_2^{(\alpha)} }^*
+ z_2^{(\alpha)} w_1^{(\alpha)}
+ z_1^{(\alpha)} w_2^{(\alpha)}
\nonumber \\
&+ {w_2^{(\alpha)}}^*
+ {z_2^{(\alpha)}}^* w_1^{(\alpha^\bullet)} 
+ \phi e^{-\imagi \theta} \Big(
z_0 z_1^{(\alpha^\bullet)} {z_1^{(\alpha)} }^* z_2^{(\alpha)} w_2^{(\alpha)}
+ z_0 z_1^{(\alpha)} z_2^{(\alpha)} {w_2^{(\alpha)}}^*
+ \imagi z_1^{(\alpha^\bullet)} {z_2^{(\alpha)}}^* 
\nonumber \\
&+ z_1^{(\alpha^\bullet)} {z_2^{(\alpha)}}^* w_1^{(\alpha)} w_2^{(\alpha)}
+ z_2^{(\alpha)} w_1^{(\alpha)} {w_2^{(\alpha)}}^*
+ \imagi z_1^{(\alpha)} |z_2^{(\alpha)}|^2
+ z_1^{(\alpha)} |w_2^{(\alpha)}|^2
+ z_1^{(\alpha)} {z_2^{(\alpha)}}^* w_1^{(\alpha^\bullet)}
\Big)
\nonumber \\
&+ \phi e^{ \imagi \theta} \Big(
z_0 z_1^{(\alpha^\bullet)} {z_2^{(\alpha)}}^* w_2^{(\alpha)}
+ z_0 z_2^{(\alpha)} {w_2^{(\alpha)} }^*
+ \imagi z_2^{(\alpha)}
+ \imagi |z_2^{(\alpha)}|^2
+ |w_2^{(\alpha)}|^2
+ {z_2^{(\alpha)}}^* w_1^{(\alpha^\bullet)} w_2^{(\alpha)}
\Big) 
\nonumber \\
&+ \phi^- e^{-\imagi \theta} \Big(
z_1^{(\alpha^\bullet)} {z_2^{(\alpha)}}^* 
+ z_1^{(\alpha)}
\Big) 
+ \phi^- e^{\imagi \theta}\Big(
z_2^{(\alpha)} + 1
\Big) 
+ e^{2\imagi \theta}\Big(
z_0 z_2^{(\alpha)} 
+ w_2^{(\alpha)}
\Big)
\nonumber \\
&+ e^{- 2\imagi \theta}\Big(
z_0 z_1^{(\alpha^\bullet)} z_1^{(\alpha)} {z_2^{(\alpha)}}^*
+ z_1^{*(\alpha^\bullet)} {z_2^{(\alpha)}}^* w_1^{(\alpha)} 
+ z_1^{*(\alpha)} {z_2^{(\alpha)}}^* w_1^{(\alpha^\bullet)} 
+ z_1^{*(\alpha)} {w_2^{(\alpha)}}^*
\Big) = 0
\nonumber \\
&\mathbf{4}. \; z_1^{(\alpha)} z_2^{(\alpha)}  w_0 
+ z_1^{(\alpha^\bullet)} {z_2^{(\alpha)}}^* w_0 
+ \imagi
+ \imagi |z_2^{(\alpha)}|^2 
+ |w_2^{(\alpha)}|^2 
+ {z_2^{(\alpha)}}^* w_1^{(\alpha^\bullet)} w_2^{(\alpha)}
\nonumber \\
&+ w_1^{(\alpha)} w_2^{(\alpha)} 
+ \phi e^{-\imagi \theta}
\Big(
z_1^{(\alpha^\bullet)} {z_1^{(\alpha)} }^* z_2^{(\alpha)} w_2^{(\alpha)} w_0
+ z_1^{(\alpha)} z_2^{(\alpha)} {w_2^{(\alpha)}}^* w_0
+ \imagi {w_2^{(\alpha)}}^* 
+ \imagi {z_2^{(\alpha)}}^* w_1^{*(\alpha^\bullet)} 
\nonumber \\
&+ \imagi |z_2^{(\alpha)}|^2 w_1^{*(\alpha)} 
+ w_1^{(\alpha)} |w_2^{(\alpha)} |^2
+ {z_2^{(\alpha)}}^* w_1^{(\alpha)}  w_1^{(\alpha^\bullet)} w_2^{(\alpha)} 
\Big) 
+  \phi e^{ \imagi \theta}
\Big(
z_1^{(\alpha^\bullet)} {z_2^{(\alpha)}}^* w_2^{(\alpha)} w_0 
+ z_2^{(\alpha)} {w_2^{(\alpha)} }^* w_0
\nonumber \\
&+ \imagi w_2^{(\alpha)}
\Big) 
+ \phi^- e^{-\imagi \theta}
\Big(
{z_2^{(\alpha)}}^* w_1^{(\alpha^\bullet)} 
+ {w_2^{(\alpha)}}^* 
+ w_1^{(\alpha)}
\Big)
+ \phi^- e^{\imagi \theta}
\Big(
w_2^{(\alpha)}
\Big)
+ e^{- 2\imagi \theta}
\Big(
z_1^{(\alpha^\bullet)} z_1^{(\alpha)} {z_2^{(\alpha)}}^* w_0 
\nonumber \\
&+ {z_2^{(\alpha)}}^*  w_1^{(\alpha^\bullet)} w_1^{(\alpha)}
+ {w_2^{(\alpha)}}^*
\Big) 
+ e^{2\imagi \theta}
\Big(z_2^{(\alpha)} w_0
\Big)=R \label{extFinalAbsRelations1}.
\end{align}
$R$ is a non-negative real number. The expression of the four power of the absolute value of extended numbers is then:
\begin{equation}
|\alpha|^4 = ( \alpha^\bullet \odot \alpha^\bullet )\cdot (\alpha \odot \alpha ) = |x|^4 + |y|^4 + R|x|^2|y|^2 \label{extAbsValueDef1}
\end{equation}
whose maps are determined by equations \ref{extFinalAbsRelations1}, which depends on the parameters $\phi$, $\theta$, and $R$.

Unfortunately, the four equations are not enough to define the maps' coefficient related to number $\alpha$, since they include coefficients of the number $\alpha^\bullet$ $z_1^{(\alpha^\bullet)}$, and $w_1^{(\alpha^\bullet)}$. We can then write the equations for the absolute value of the number $\alpha^\bullet$:
\begin{equation}
|\alpha^\bullet|^4 = ( (\alpha^\bullet)^\bullet \odot (\alpha^\bullet)^\bullet )\cdot (\alpha^\bullet \odot \alpha^\bullet ) = |x_{\alpha^\bullet}|^4 + |y_{\alpha^\bullet}|^4 + R|x_{\alpha^\bullet}|^2|y_{\alpha^\bullet}|^2.
\end{equation}
We will obtain a similar set of equations like \ref{extFinalAbsRelations1} replacing the complex numbers $x,y$ by the extended and imaginary parts of the number $\alpha^\bullet$ and the coefficients by $z_1^{(\alpha^\bullet)},w_1^{(\alpha^\bullet)}$ and $z_2^{(\alpha^\bullet)}, w_2^{(\alpha^\bullet)}$. However, we still have coefficients related to the number $(\alpha^\bullet)^\bullet$. One way to solve this issue is to impose a closure condition. In this case, we can propose the condition
\begin{equation}
(\alpha^\bullet)^\bullet = \alpha, \label{closureCondition_0}
\end{equation}
or, using the extended and complex parts:
\begin{align}
\mathbf{5}.\; &z_2^{(\alpha)} w_2^{(\alpha^\bullet)} + w_2^{(\alpha)} =0 \nonumber \\
\mathbf{6}.\; &z_2^{(\alpha)} z_2^{(\alpha^\bullet)} - 1 =0 \label{closureCondition}
\end{align}

In that case, we have
\begin{equation}
|\alpha^\bullet|^4  =  ( (\alpha^\bullet)^\bullet \odot (\alpha^\bullet)^\bullet )\cdot (\alpha^\bullet \odot \alpha^\bullet ) = (\alpha \odot \alpha) \cdot ( \alpha^\bullet \odot \alpha^\bullet ) = |\alpha|^4.
\end{equation}
The third and fourth equations of the set of equations \ref{extFinalAbsRelations1} are reduced to the third and fourth equations of equations \ref{extFinalAbsRelations1}. The first and second equations of the equations for the absolute value of the number $\alpha^\bullet$, obtained from equations \ref{extConjDangerous} to the product $ (\alpha^\bullet)^\bullet \odot (\alpha^\bullet)^\bullet $:
\begin{align}
z_1^{({\alpha^\bullet}^\bullet)} {z_2^{(\alpha^\bullet)}}^* w_2^{(\alpha^\bullet)} + z_2^{(\alpha^\bullet)} {w_2^{(\alpha^\bullet)}}^* &= 0
\nonumber \\
-|z_2^{(\alpha^\bullet)}|^2 
+ \imagi {z_2^{(\alpha^\bullet)}}^* w_1^{({\alpha^\bullet}^\bullet)} w_2^{(\alpha^\bullet)} 
+ \imagi|w_2^{(\alpha^\bullet)}|^2 &= 1
\end{align}
are modified using $z_1^{({\alpha^\bullet}^\bullet)} = z_1^{(\alpha)}$ and $w_1^{({\alpha^\bullet}^\bullet)} = w_1^{(\alpha)}$ according to the closure condition, like
\begin{align}
z_1^{(\alpha)} {z_2^{(\alpha^\bullet)}}^* w_2^{(\alpha^\bullet)} + z_2^{(\alpha^\bullet)} {w_2^{(\alpha^\bullet)}}^* &= 0
\nonumber \\
-|z_2^{(\alpha^\bullet)}|^2 
+ \imagi {z_2^{(\alpha^\bullet)}}^* w_1^{(\alpha)} w_2^{(\alpha^\bullet)} 
+ \imagi|w_2^{(\alpha^\bullet)}|^2 &= 1 \label{extConjRelations_1}.
\end{align}

The previous equations are not independent of the set of equations \ref{extFinalAbsRelations1} and  \ref{closureCondition} because the result of applying the closure condition on the absolute value expression for the number $\alpha^{\bullet}$ is the same as for $\alpha$. Nevertheless, it explicitly shows the dependency of the equation system of the quantities $ z_2^{(\alpha^\bullet)}$ and $ w_2^{(\alpha^\bullet)}$.

According to the above results, we have eight variables: $z_1^{(\alpha)},w_1^{(\alpha)}, z_2^{(\alpha)}, w_2^{(\alpha)}, z_1^{(\alpha^\bullet)},w_1^{(\alpha^\bullet)}, z_2^{(\alpha^\bullet)}$, and $ w_2^{(\alpha^\bullet)}$ and six equations. That means we are two complex or one extended equation short. These equations will be added once the property of the Conjugate symmetry are discussed below.


\subsection{The R-parameter}

The set of extended complex numbers $\mathbb{E}$ can be written as $a + ib + \extk c +i \extk d $, where  $a,b,c,d \in \mathbb{R}$. The parameter $R$ can be proposed in analogy with the isotropic property for linear spaces. The isotropic property states that the length of a vector remains invariant under an axis rotation. If we look at the complex vector space, all numbers laying on the same centered circle, such as the complex number $z=a + \imagi b$, have the same absolute value, equal to $|z| = \sqrt{a^2 + b^2}$. Some of these numbers lying on the sphere are $z_1 = a + \imagi b$, $z_2 = -a + \imagi b$, $z_3 = a - \imagi b$, $z_4 = -a - \imagi b$, $z_5 = b + \imagi a$, $z_6 = -b + \imagi a$, $z_7 = b - \imagi a$ and $z_8 = -b - \imagi a$. 

The absolute value of extended number $\alpha = a + \imagi b + \extk c + \imagi \extk d$, where $a,b,c,d \in \mathbb{R}$ must, then, remain constant when the axis is rotated, which means that numbers:
\begin{align*}
\alpha_1 = \pm a \pm \imagi b \pm  \extk c \pm  \imagi \extk d \\
\alpha_2 = \pm b \pm \imagi c \pm  \extk d \pm  \imagi \extk a \\
\alpha_3 = \pm c \pm \imagi d \pm  \extk a \pm  \imagi \extk b \\
\alpha_4 = \pm d \pm \imagi a \pm  \extk b \pm  \imagi \extk c 
\end{align*}
should have the same absolute value as $\alpha$.

The proposed absolute value for an extended number raised to the fourth power, as shown in equations  \ref{extAbsValueDef1}, is
\begin{equation}
|\alpha|^4  = |x|^4 + |y|^4 + R|x|^2|y|^2 \quad x,y \in  \mathbb{C}\quad \text{and} \quad R \in \mathbb{R}, R\geq 0
\end{equation}
being $x$ and $y$ the extended and imaginary part of the extended number. If we substitute $ x = a \imagi + b $ and $y =c \imagi +d$, being $a,b,c,d \in \mathbb{R}$, we have
\begin{align}
|\alpha|^4 =& (a^2+b^2)^2+ (c^2+d^2)^2 + R(a^2+b^2)(c^2+d^2)
\nonumber \\
=& a^4 + b^4 + c^4 + d^4 + 2a^2b^2  + 2c^2d^2 + R(a^2c^2 + a^2d^2 + b^2c^2 + b^2d^2).
\end{align}
$R=2$ is the only possible value for the absolute value 
\begin{equation}
|\alpha|^4 = a^4 + b^4 + c^4 + d^4 + 2( a^2b^2  + a^2c^2 + a^2d^2 + b^2c^2 + b^2d^2  + c^2d^2),
\end{equation}
remain constant under any permutation of $a,b,c,d$ with any combination of $\pm$ sign.

The final form for the absolute value of an extended number $\alpha = x \extk + y$ is then
\begin{equation}
|\alpha|^4  = |x|^4 + |y|^4 + 2|x|^2|y|^2.
\end{equation}

\section{Division between extended numbers}

In the complex number domain, the division of two complex numbers can be accomplished by multiplying the numerator and denominator by the complex conjugate of the denominator. To do a similar procedure for the division between extended numbers, we must first review the existence of extraneous and missing solutions when the same factor multiplies both members of an extended equation. The extraneous solution (or spurious solution) emerges from solving the problem while a missing solution is a valid solution that of the original problem, but disappeared along with the solution. We study the necessity and sufficiency for the equality of an extended equation, before and after the multiplication. 

Let us consider the extended equation
\begin{equation}
\alpha=\beta \quad \forall \;\alpha,\beta \; \in \mathbb{E}. \label{extSameFactorMultEq}
\end{equation}
If we multiply both members of the equation \ref{extSameFactorMultEq} by a third extended factor $\gamma$, will the equality of the new equation still hold? Moreover, if it does, would it introduce or eliminate solutions to the original equations? 

Let us analyze different cases:
\begin{itemize}
\item The standard multiplication of both members by an extended number
\begin{equation}
\alpha \cdot \gamma \stackrel{?}{=} \beta \cdot \gamma \quad \forall \; \alpha,\beta, \gamma \; \in \mathbb{E}
\label{extSameFactorMultEq1}
\end{equation}
The equation $\alpha=\beta$ means that their extended and imaginary parts are equals, respectively, like $\alpha_E=\beta_E $ and $\alpha_I=\beta_I $. The members of the equation \ref{extSameFactorMultEq1} have the form:
\begin{align}
\alpha \cdot \gamma = \big( \alpha_E \gamma_E z_0 + \alpha_E \gamma_I + \alpha_I \gamma_E \big)\extk + \alpha_E \gamma_E w_0 + \alpha_I \gamma_I
\nonumber \\
\beta \cdot \gamma = \big( \beta_E \gamma_E z_0 + \beta_E \gamma_I + \beta_I \gamma_E \big)\extk + \beta_E \gamma_E w_0 + \beta_I \gamma_I,
\end{align}
which led to the equations
\begin{align}
( \alpha_E - \beta_E )( \gamma_E z_0 + \gamma_I ) + ( \alpha_I - \beta_I) \gamma_E = 0
\nonumber \\
( \alpha_E - \beta_E ) \gamma_E w_0 + ( \alpha_I - \beta_I) \gamma_I = 0. \label{extSameFactorMultEq2}
\end{align}
If $\alpha_E=\beta_E $ and $\alpha_I=\beta_I $, the sufficiency of the statement is probed. However, the necessity for the inverted proposition is not true. Indeed, from equations \ref{extSameFactorMultEq2}, we cannot extract the initial equality $\alpha=\beta$, and that is because of the appearances of extraneous solutions. Multiplying first complex equation of \ref{extSameFactorMultEq2} by $\gamma_I$, the second by $\gamma_E$ and subtract one equation from another we obtain:
\begin{equation}
( \alpha_E - \beta_E )( \gamma_I^2 + \gamma_E \gamma_I z_0 - \gamma_E^2 w_0)=0,
\end{equation}
which indicate the presence of a new solution related to the case 
\begin{equation}
\gamma_I^2 + \gamma_E \gamma_I z_0 - \gamma_E^2 w_0=0.
\end{equation}
The necessity can be easily probed for some specific cases like $(\forall \; \gamma_E=0, \gamma_I\neq 0)$ or $(\forall \; \gamma_E\neq 0, \gamma_I= 0, w_0 \neq 0 )$.

\item  The left conjugated multiplication of an extended number, 
\begin{equation}
\gamma \odot \alpha \stackrel{?}{=} \gamma \odot \beta, 
\end{equation}
the equation for every member is
\begin{align}
\gamma \odot \alpha = \big( \gamma_E^* \alpha_I z_1^{(\gamma)} + \gamma_I^* \alpha_E  \big)\extk + \gamma_E^* \alpha_I w_1^{(\gamma)} + \imagi \gamma_E^* \alpha_E + \gamma_I^* \alpha_I
\nonumber \\
\gamma \odot  \beta = \big( \gamma_E^* \beta_I z_1^{(\gamma)} + \gamma_I^* \beta_E  \big)\extk + \gamma_E^* \beta_I w_1^{(\gamma)} + \imagi \gamma_E^* \beta_E + \gamma_I^* \beta_I.
\end{align}
Grouping the extended and imaginary part and setting equal to zero, we have
\begin{align}
\gamma_E^*  z_1^{(\gamma)} (\alpha_I - \beta_I)  + \gamma_I^* (\alpha_E - \beta_E) = 0
\nonumber \\
(\gamma_E^* w_1^{(\gamma)} + \gamma_I^*) ( \alpha_I - \beta_I ) + \imagi \gamma_E^* (\alpha_E - \beta_E ) = 0  \label{extSameFactorMultEq3}
\end{align}
which probe the sufficiency of the statement. After multiplying the first complex equation by Equation \ref{extSameFactorMultEq3}, the second by $\gamma_I^*$ and subtract one equation from another, head to the equation
\begin{equation}
\big[ (\gamma_I^*)^2 + \gamma_E^* \gamma_I^* w_1^{(\gamma)} - \imagi (\gamma_E^*)^2  z_1^{(\gamma)} \big ] ( \alpha_I - \beta_I ) = 0
\end{equation}
which shows the inclusion of a new solution from the extended number $\gamma$ that satisfy the equation
\begin{equation}
(\gamma_I^*)^2 + \gamma_E^* \gamma_I^* w_1^{(\gamma)} - \imagi (\gamma_E^*)^2  z_1^{(\gamma)} =0
\end{equation}

\item   The right conjugated multiplication of an extended number, 
\begin{equation}
\alpha \odot \gamma \stackrel{?}{=} \beta \odot \gamma , 
\end{equation}
led to equations:
\begin{align}
(\alpha_E^* z_1^{(\alpha)} - \beta_E^*  z_1^{(\beta)})\gamma_I  + ( \alpha_I^* - \beta_I^* )\gamma_E = 0
\nonumber \\
(\alpha_E^* w_1^{(\alpha)} - \beta_E^*  w_1^{(\beta)} +  \alpha_I^* - \beta_I^*)  \gamma_I + \imagi (\alpha_E^* - \beta_E^* )\gamma_E  = 0.  \label{extSameFactorMultEq4}
\end{align}

In this case, the sufficiency of the statement can be verified because if numbers $\alpha,\beta$ are equals, then also their maps. We can not verify the necessity for this statement due to the impossibility of factorizing the terms with $z_1$ and $w_1$.
\end{itemize}

We can now define the division between extended numbers. Our first proposal is similar to the division between complex numbers, where the numerator and denominator are multiplied by the extended complex conjugates of the denominator. Follow this reasoning; our initial proposition is first, conjugately left multiply both numerator and denominator by the number like $ (\alpha \odot \alpha)$ and then standard multiply by the expression $\alpha^{\bullet}  \odot \alpha^{\bullet}$ like
\begin{equation}
\frac{\lambda}{\alpha} = \frac{(\alpha \odot \lambda) }{ (\alpha \odot \alpha)} 
= \frac{(\alpha \odot \lambda) \cdot (\alpha^{\bullet}  \odot \alpha^{\bullet} )}{ (\alpha \odot \alpha) \cdot (\alpha^{\bullet}  \odot \alpha^{\bullet} )}
 = \frac{(\alpha \odot \lambda)\cdot (\alpha^{\bullet}  \odot \alpha^{\bullet} )}{|\alpha|^4}.
\end{equation}
This proposal is incorrect because we cannot obtain $\alpha$ with an inverse process from the number resulting from the division. That means that we cannot get the above number by any multiplication:
\begin{equation}
\frac{\lambda}{\alpha} \odot \alpha = \alpha \odot \frac{(\alpha \odot \lambda)\cdot (\alpha^{\bullet}  \odot \alpha^{\bullet} )}{|\alpha|^4} \neq \lambda 
\qquad \text{ or } \qquad 
\frac{\lambda}{\alpha} \cdot \alpha = \alpha \cdot \frac{(\alpha \odot \lambda)\cdot (\alpha^{\bullet}  \odot \alpha^{\bullet} )}{|\alpha|^4} \neq \lambda
\end{equation}
We propose the standard multiplication of both numerator and denominator by the expression $(\sqrt[*]{\alpha})^\bullet \odot (\sqrt[*]{\alpha})^\bullet$, where we remember that the conjugated root is the inverse operation of the conjugated product. Indeed, any extended number can be expressed as the conjugated product of its conjugated root, like $\alpha = \sqrt[*]{\alpha} \odot \sqrt[*]{\alpha}$. The new proposal for the division between extended numbers is then:
\begin{equation}
\frac{\lambda}{\alpha} 
\equiv \frac{ \lambda}{ (\sqrt[*]{\alpha} \odot \sqrt[*]{\alpha})} 
= \frac{\lambda \cdot  [(\sqrt[*]{\alpha})^\bullet \odot (\sqrt[*]{\alpha})^\bullet]}
	{ { [\sqrt[*]{\alpha} \odot \sqrt[*]{\alpha}]} \cdot { [(\sqrt[*]{\alpha})^\bullet \odot (\sqrt[*]		{\alpha})^\bullet]}} 
= \frac{\lambda \cdot  [(\sqrt[*]{\alpha})^\bullet \odot (\sqrt[*]{\alpha})^\bullet]}{|\sqrt[*]{\alpha}|^4}.
\end{equation}
Now we can verify that:
\begin{equation}
\frac{\lambda}{\alpha} \cdot \alpha =\alpha \cdot \frac{\lambda \cdot  [(\sqrt[*]{\alpha})^\bullet \odot (\sqrt[*]{\alpha})^\bullet]}{|\sqrt[*]{\alpha}|^4} 
=  \frac{\lambda (\sqrt[*]{\alpha} \odot \sqrt[*]{\alpha}) \cdot  [(\sqrt[*]{\alpha})^\bullet \odot (\sqrt[*]{\alpha})^\bullet]}{|\sqrt[*]{\alpha}|^4} 
= \frac{\lambda |\sqrt[*]{\alpha}|^4}{|\sqrt[*]{\alpha}|^4} = \lambda
\end{equation}

For $\lambda=1$, we can probe the existence of multiplicative inverse $\alpha^-$ as one of the properties that the extended numbers satisfy. The multiplicative inverse has the form
\begin{equation}
\alpha^- =\frac{1}{\alpha} = \frac{ 1}{ (\sqrt[*]{\alpha} \odot \sqrt[*]{\alpha})} 
= \frac{[(\sqrt[*]{\alpha})^\bullet \odot (\sqrt[*]{\alpha})^\bullet]}
	{ { [\sqrt[*]{\alpha} \odot \sqrt[*]{\alpha}]} \cdot { [(\sqrt[*]{\alpha})^\bullet \odot (\sqrt[*]		{\alpha})^\bullet]}} 
= \frac{[(\sqrt[*]{\alpha})^\bullet \odot (\sqrt[*]{\alpha})^\bullet]}{|\sqrt[*]{\alpha}|^4}
\end{equation}

There is another possibility for the division between extended numbers since we can also define the extended conjugated root $\sqrt[\bullet]{\alpha}$ as the inverse operation of the conjugated product of the $()^\bullet$-map like
\begin{equation}
\alpha = (\sqrt[\bullet]{\alpha})^\bullet \odot (\sqrt[\bullet]{\alpha})^\bullet.
\end{equation}
In this case, we can propose other division like
\begin{equation}
\frac{\lambda}{\alpha} 
= \frac{ \lambda}{ (\sqrt[\bullet]{\alpha})^\bullet \odot (\sqrt[\bullet]{\alpha})^\bullet} 
= \frac{\lambda \cdot  [\sqrt[\bullet]{\alpha} \odot \sqrt[\bullet]{\alpha}]}
	{ [\sqrt[\bullet]{\alpha} \odot \sqrt[\bullet]{\alpha}] \cdot { [(\sqrt[\bullet]{\alpha})^\bullet \odot (\sqrt[\bullet]		{\alpha})^\bullet]}} 
= \frac{\lambda \cdot  [\sqrt[\bullet]{\alpha} \odot \sqrt[\bullet]{\alpha}]}{|\sqrt[\bullet]{\alpha}|^4}
\end{equation}
and the inverse like
\begin{equation}
\alpha^- =\frac{[\sqrt[\bullet]{\alpha} \odot \sqrt[\bullet]{\alpha}]}{|\sqrt[\bullet]{\alpha}|^4}
\end{equation}
Rest now to interpret the double existence of two divisions, and with it, the existence of two inverse for every extended number.

\section{Extended domain properties. Absolute value.} \label{extPropSection}

The dependency of the maps $\extk^2$, $\extk^*$, and $\extk^\bullet$ on parameters $\phi$, $\theta$ and $R$, provide the extended domain with some properties:
\begin{itemize}
\item 
The absolute value raised to the fourth power of an extended number is:
\begin{equation}
\langle \alpha  , \alpha, \alpha , \alpha \rangle  = |x|^4 + |y|^4 + 2|x|^2|y|^2 \qquad x,y \in \mathbb{C},
\end{equation}
and the absolute value is defined as:
\begin{equation}
|\alpha| = \sqrt[4]{|x|^4 + |y|^4 + 2|x|^2|y|^2 }.
\end{equation}

\item 
The absolute value of an extended number is zero if the extended and the imaginary part of the number are also zero.
\item 
It is straightforward to prove the property of the extended numbers that:
\begin{align*}
(c\alpha)^* = c^*\alpha^* \qquad \text{ and } \qquad (c\alpha)^\bullet = c\alpha^\bullet 
\qquad \forall \; c \in \mathbb{C}, \alpha \in \mathbb{E}.
\end{align*}

\item 
The relations for obtaining the extended and conjugated maps for pure extended and pure complex numbers result in undetermined values for numbers $z_1,z_2,w_1$ and $w_2$. That is because of the dependency of equations \ref{extFinalParam} on the quantities $\phi$ and $\phi^-$, whose values turn zero or infinity for $|x|$ and/or $|y|$ being zero. However, both cases are trivial, since taking $\alpha = x \extk + y$, for $y=0$ we have
\begin{equation}
x\extk \odot x\extk=|x|^2i \qquad \text{and} \qquad [(x\extk)^\bullet \odot (x\extk)^\bullet][(x\extk) \odot (x\extk)] = |x|^4,
\end{equation}
while for $x=0$, we have the typical operations between complex numbers.

\item 
The extended numbers $\alpha \equiv x \extk + y$ and $c\alpha = c(x \extk + y)\;\; c \in \mathbb{C}$ have equaled values of $\theta$ and $\phi$. From their definitions, we have:  
\begin{equation}
\phi_{c\alpha} = \frac{|cx|}{|cy|} \frac{|c||x|}{|c||y|} = \frac{|x|}{|y|}\equiv \phi_{\alpha} \qquad \theta_{c\alpha} = (\theta_x+\theta_c) - (\theta_y + \theta_c) =  \theta_x - \theta_y \equiv \theta_{\alpha}. \label{extThetaPhiInvariant}
\end{equation}

\item 
From the expression of the absolute value in Eq. \ref{extAbsValueDef1} and the invariant character of $\theta$ and $\phi$ for extended numbers number $\alpha$ and $c\alpha$ where $c \in \mathbb{C}$, as shown in Eq. \ref{extThetaPhiInvariant}, the absolute value of $c\alpha$ have the expression:
\begin{align}
|c\alpha| =& \sqrt[4]{[(c\alpha^{\bullet})\odot( c\alpha^{\bullet} )] \cdot [(c\alpha \odot c\alpha)]} 
\nonumber \\
 =&\sqrt[4]{c^*c\;c^*c[(\alpha^{\bullet}  \odot \alpha^{\bullet} \cdot (\alpha \odot \alpha)]}
\nonumber \\
=& \sqrt[4]{|c|^4[(\alpha^{\bullet}  \odot \alpha^{\bullet} \cdot (\alpha \odot \alpha)]} 
\nonumber \\
=& |c||\alpha|
\end{align}
\end{itemize}
\section{Linearity}

Linearity is another property that the inner product must satisfy since it is the base property of the superposition principle. From the algebraic properties of the extended numbers, specifically those related to the standard product, we found that the complex products $\alpha \odot \beta$ satisfy the distribution and associative property like shown on equation \ref{ExtVectorDistrProp}:
\begin{equation*}
(\alpha_1 \odot \beta_1)\cdot \big[(\alpha_2 \odot \beta_2) + (\alpha_3 \odot \beta_3)\big] = (\alpha_1 \odot \beta_1)\cdot(\alpha_2 \odot \beta_2) + (\alpha_1 \odot \beta_1)\cdot(\alpha_3 \odot \beta_3).
\end{equation*}
and products like Eq.$(\alpha_1^\bullet \odot \beta_1^\bullet)\cdot (\alpha_2 \odot \beta_2)$
satisfies the distributive properties of equation \ref{ExtProdDistrProp1}
\begin{equation*}
(\alpha_1^\bullet \odot \beta_1^\bullet)\cdot \big[(\alpha_2 \odot \beta_2) + (\alpha_3 \odot \beta_3)\big] = (\alpha_1^\bullet \odot \beta_1^\bullet) \cdot (\alpha_2 \odot \beta_2) + (\alpha_1^\bullet \odot \beta_1^\bullet) \cdot (\alpha_3 \odot \beta_3)
\end{equation*}

Even when the two-dimensional vector satisfies Linearity's property, we also study what conditions the addends of the factors of the inner product should satisfy for the product remain linear. 

For simplicity, we express the inner product of the equation \ref{ExtInnerProdDef} in the form of a two-row matrix
\begin{equation*}
\extproduct{\alpha}{\beta}{\gamma}{\delta}
\equiv (\alpha^{\bullet}  \odot \beta^{\bullet}) \cdot (\gamma \odot \delta).
\end{equation*}

The Linearity Property of the internal factors implies that:
\begin{align}
\extproduct{\alpha_1 + \alpha_2}{\beta_1 + \beta_2}{\gamma}{\delta}  &=
\extproduct{\alpha_1}{\beta_1}{\gamma}{\delta} +  \extproduct{\alpha_2}{\beta_2}{\gamma}{\delta} 
\nonumber \\
\extproduct{\alpha}{\beta}{\gamma_1 + \gamma_2}{\delta_1 + \delta_2}  &=
\extproduct{\alpha}{\beta}{\gamma_1}{\delta_1} + \extproduct{\alpha}{\beta}{\gamma_2}{\delta_2}
 \label{extLinearitySum}
\end{align}
and also
\begin{align}
\extproduct{\alpha_1 \cdot \alpha_2}{\beta_1 \cdot \beta_2}{\gamma}{\delta}  &=
(\alpha_1^\bullet \odot \beta_1^\bullet)  \extproduct{\alpha_2}{\beta_2}{\gamma}{\delta} 
= (\alpha_2^\bullet \odot \beta_2^\bullet)  \extproduct{\alpha_1}{\beta_1}{\gamma}{\delta}. 
\nonumber \\
\extproduct{\alpha}{\beta}{\gamma_1 \cdot \gamma_2}{\delta_1 \cdot \delta_2}  &=
(\gamma_1 \odot \delta_1 ) \; \extproduct{\alpha}{\beta}{\gamma_2}{\delta_2}
 = (\gamma_2 \odot \delta_2 ) \; \extproduct{\alpha}{\beta}{\gamma_1}{\delta_1}.
\label{extLinearityEscalar}
\end{align}

The conditions for the Linearity of inner products of equations \ref{extLinearitySum} and \ref{extLinearityEscalar} being satisfied can be computed using the algebraic properties of the operations defined above. We do not analyze the Linearity for the internal factors since it is out of the scope of this work, even this property can be the base for the explanation of phenomena using a theory developed on extended numbers. We point out, nevertheless, the existence of two types of Properties of Linearity and also that the proposed inner product satisfied the Linearity for the extended product pair. That is essential to ensure the superposition principle for the quantum state vectors for $n$-VMVF systems.


\section{Conjugate symmetry} \label{ConjugateSymmetry}

The main objective of this works to propose a new Lorentz space where a new Quantum Theory for $n$-VMVF systems can be developed.  That is the guide when proposing new properties of the studied space. Indeed, from the classical theory obtained for $n$-VMVF systems, we proposed the extended unit definition and also an inner product that satisfies the superposition principle for a pair of extended numbers. We proposed that the conjugated product of two extended numbers as a representation of the quantum state of the system. The necessity of including the ``\textit{bra}'' state, demands the vector space to have a symmetric bilinear form. For that reason, we need to impose the Conjugate symmetry axiom for the pairs composing the inner product.

In the space of complex vectors, the property of the conjugate symmetry has the form:
\begin{equation}
\langle u,v \rangle = \overline{\langle v,u \rangle} \qquad \forall \; u,v \in \mathbb{C}
\end{equation}
where $\overline{()}$ is referred to as the map $J: V \to V^*$. Is very well known that the  conjugate symmetry ensures that the complex inner product that satisfies:
\begin{equation*}
\langle u,v + w\rangle  = \langle u,v \rangle +\langle u, w\rangle\; \text{and } \;
\langle u,w v \rangle   = w \langle u,v \rangle  
\end{equation*}
also, satisfy
\begin{equation*}
\langle u + w,v \rangle  = \langle u,v \rangle +\langle w, v\rangle\; \text{and } \;
\langle w u, v \rangle   = w^* \langle u,v \rangle  
\end{equation*}

In the extended case, the expression for the inner product 
\begin{equation*}
\extproduct{\alpha}{\beta}{\gamma}{\delta}
\equiv (\alpha^{\bullet}  \odot \beta^{\bullet}) \cdot (\gamma \odot \delta),
\end{equation*}
show us the existence of the map $J: V \to V', \; V,V' \in \mathbb{E} $ that  $J (\alpha \odot \beta) \to (\alpha^\bullet \odot \beta^\bullet)$. We refer to this map as the extended pair conjugate and denote it as $J \equiv ()^\cbullet$.

\subsection{The $()^\cbullet$ map}
Unlike the complex numbers, the properties of the map $J: V \to V'$ can not be extracted from the positive-definiteness axiom or the coefficients $z_1^{(\alpha)},w_1^{(\alpha)}, z_2^{(\alpha)}, w_2^{(\alpha)}, z_1^{(\alpha^\bullet)},w_1^{(\alpha^\bullet)}, z_2^{(\alpha^\bullet)}$, and $ w_2^{(\alpha^\bullet)}$. Instead, with no loss of generality, we can impose the properties to this map and relate it then with the above coefficients. The properties that this map must satisfy are 
\begin{itemize}
\item Uniqueness. Being the extended pair conjugate an operation applied on a complex product, it must ensure that the map is unique; otherwise, we will have conflicts for a number expressed as different complex products.
\item Associativity for the Sum. The bi-linear form of the inner product demands that:
\begin{equation}
(\alpha + \beta)^\cbullet = (\alpha)^\cbullet + (\beta)^\cbullet \label{ext_pair_conj_axiom_sum}
\end{equation}
\item Associativity for the Standard Product. Same as before, the bi-linear form of the inner product demands that:
\begin{equation}
(\alpha \cdot \beta)^\cbullet = (\alpha)^\cbullet \cdot (\beta)^\cbullet \label{ext_pair_conj_axiom_prod}
\end{equation}
\end{itemize}

As the expressions for the parameters $z_i^{(\alpha)},w_i^{(\alpha)}$ are unknown, the map's uniqueness can be established by imposing the map being a function of the extended number resulting from the extended product, $e.i.$:
\begin{equation}
\text{if}\;\; \alpha \odot \beta = \gamma \; \text{then }\; (\alpha \odot \beta)^\cbullet= \alpha^\bullet \odot \beta^\bullet \equiv \mathcal{F}(\gamma) \equiv \mathcal{F}_E(\gamma_E, \gamma_I) \extk + \mathcal{F}_I(\gamma_E, \gamma_I),
\end{equation}
being $\gamma_E, \gamma_I$, the extended and the imaginary part of the extended number $\gamma$ and $\mathcal{F}_E, \mathcal{F}_I$ the same quantities for the extended function $\mathcal{F}$.

We can also impose that the last two properties of associativity for the function $\mathcal{F}(\gamma)$:
\begin{align}
\mathcal{F}(\alpha) + \mathcal{F}(\beta)  &= \mathcal{F}(\alpha + \beta) \\
\mathcal{F}(\alpha) \cdot \mathcal{F}(\beta)  &= \mathcal{F}(\alpha \cdot \beta) 
\end{align}

The Associativity for the Sum is satisfied if the proposed function is linear. Let us express function $\mathcal{F}(\alpha)$ with the linear expression:
\begin{equation}
\mathcal{F}(\alpha) = \mathcal{F}_E(\alpha_E, \alpha_I) \extk + \mathcal{F}_I(\alpha_E, \alpha_I) \equiv (a_1 \alpha_E +  a_2 \alpha_I)\extk + a_3 \alpha_E +  a_4 \alpha_I, \label{F_linear_representation}
\end{equation}
where $a_1, a_2, a_3, a_4$ are complex numbers. Replacing this expression on the sum of the maps of the extended numbers $\alpha$ and $\beta$ we obtain:
\begin{align}
\mathcal{F}(\alpha)  + \mathcal{F}(\beta)  =& 
(a_1 \alpha_E +  a_2 \alpha_I)\extk + a_3 \alpha_E +  a_4 \alpha_I
+ (a_1 \beta_E +  a_2 \beta_I)\extk + a_3 \beta_E +  a_4 \beta_I
\nonumber \\
=& [ a_1 (\alpha_E + \beta_E) +  a_2 (\alpha_I + \beta_I )]\extk + a_3 (\alpha_E + \beta_E) +  a_4 (\alpha_I + \beta_I )
\nonumber \\
=& \mathcal{F}(\alpha + \beta),
\end{align}
which proves the Associativity for the Sum for the function \ref{F_linear_representation}. 

The Associativity for the Standard Product can not be proved because the coefficients for the standard product are not defined. However, we can obtain their values from this property. From the properties of $()^\cbullet$ map, the Associativity for the Standard Product, equation \ref{ext_pair_conj_axiom_prod}, demands that:
\begin{equation*}
 \mathcal{F}(\alpha \cdot \beta) =  \mathcal{F}(\alpha ) \cdot  \mathcal{F}(\beta).
\end{equation*}
Replacing the linear expressions \ref{F_linear_representation}, we can find relations between the coefficients $a_1, a_2, a_3, a_4$ and $z_0, w_0$. Indeed, the extended and the imaginary part of the expression
\begin{align}
 \mathcal{F}(\alpha \cdot \beta)  = \mathcal{F}\big( (z_0 \alpha_E \beta_E + \alpha_E \beta_I+ \alpha_I \beta_E )\extk + w_0 \alpha_E \beta_E  +\alpha_I \beta_I  \big)
 \end{align}
 are
\begin{align}
 \mathcal{F}_E(\alpha \cdot \beta) =& a_1(z_0 \alpha_E \beta_E + \alpha_E \beta_I+ \alpha_I \beta_E) + a_2(w_0 \alpha_E \beta_E  +\alpha_I \beta_I)
  \nonumber \\
 \mathcal{F}_I(\alpha \cdot \beta) =& a_3(z_0 \alpha_E \beta_E + \alpha_E \beta_I+ \alpha_I \beta_E) + a_4(w_0 \alpha_E \beta_E  +\alpha_I \beta_I). \label{ext_pair_conj_axiom_prod_MI}
\end{align}
The extended and the imaginary part of the expression
\begin{align}
 \mathcal{F}(\alpha) \cdot \mathcal{F}(\beta)  =& [z_0 \mathcal{F}_E(\alpha) \mathcal{F}_E(\beta) + \mathcal{F}_E(\alpha) \mathcal{F}_I(\beta)+ \mathcal{F}_I(\alpha) \mathcal{F}_E(\beta)] \extk 
  \nonumber \\
 &+  w_0 \mathcal{F}_E(\alpha) \mathcal{F}_E(\beta)  + \mathcal{F}_I(\alpha) \mathcal{F}_I(\beta)
 \end{align}
 are:
\begin{align} 
 [\mathcal{F}(\alpha) \cdot \mathcal{F}(\beta)]_E =& 
 z_0 (a_1 \alpha_E +  a_2 \alpha_I)(a_1 \beta +  a_2 \beta_I)
  \nonumber \\
 & + (a_1 \alpha_E +  a_2 \alpha_I)(a_3 \beta +  a_4 \beta_I)
  \nonumber \\
 & + (a_3 \alpha_E +  a_4 \alpha_I)(a_1 \beta +  a_2 \beta_I)
  \\
 [\mathcal{F}(\alpha) \cdot \mathcal{F}(\beta)]_I =& 
 w_0 (a_1 \alpha_E +  a_2 \alpha_I)(a_1 \beta +  a_2 \beta_I)
  \nonumber \\
 & + (a_3 \alpha_E +  a_4 \alpha_I)(a_3 \beta +  a_4 \beta_I) \label{ext_pair_conj_axiom_prod_MD}
\end{align}
Expanding the equations \ref{ext_pair_conj_axiom_prod_MI} and \ref{ext_pair_conj_axiom_prod_MD}, and setting equals its extended and imaginary parts, we obtain:
\begin{align}
&\alpha_E \beta_E (z_0 a_1^2 + 2 a_1 a_3 - z_0 a_1 - w_0 a_2)
\nonumber \\
 + & \alpha_E \beta_I(z_0 a_1 a_2 + a_1 a_4 + a_3 a_2 - a_1)
\nonumber \\
 + &  \alpha_I \beta_E( z_0 a_2 a_1 + a_2 a_3 + a_4 a_1 - a_1)
\nonumber \\
 + &  \alpha_I \beta_I (z_0 a_2^2 + 2 a_2 a_4 - a_2) = 0
\end{align}
and
\begin{align}
&\alpha_E \beta_E (w_0 a_1^2 + a_3^2 - z_0 a_3 - w_0 a_4)
\nonumber \\
 + & \alpha_E \beta_I(w_0 a_1 a_2 + a_3 a_4 - a_3)
\nonumber \\
 + &  \alpha_I \beta_E( w_0 a_2 a_1 + a_4 a_3 - a_3)
\nonumber \\
 + &  \alpha_I \beta_I (w_0 a_2^2 + a_4^2 - a_4) = 0.
\end{align}	
Being $\alpha$ and $\beta$ any two extended numbers, this equation holds if all the coefficients of $\alpha_i, \beta_j$ are zero. Note that $\alpha_E \beta_I$ and $\alpha_I \beta_E$ have the same coefficient for the extended and imaginary part, respectively. The independent set of equations for the equation \ref{ext_pair_conj_axiom_prod} hold for any pair of extended numbers is
\begin{align}
z_0 a_1^2 + 2 a_1 a_3 - z_0 a_1 - w_0 a_2 &= 0
\nonumber \\
z_0 a_1 a_2 + a_1 a_4 + a_2 a_3 - a_1 &= 0
\nonumber \\
z_0 a_2^2 + 2 a_2 a_4 - a_2  &= 0
\nonumber \\
w_0 a_1^2 + a_3^2 - z_0 a_3 - w_0 a_4  &= 0
\nonumber \\
w_0 a_1 a_2 + a_3 a_4 - a_3  &= 0
\nonumber \\
w_0 a_2^2 + a_4^2 - a_4  &= 0 \label{ext_pair_conj_axiom_Coeff_equations}
\end{align}
We obtain six equations for the same numbers of variables: $a_1, a_2, a_3, a_4$ and $z_0, w_0$. This result shows us that we can indeed express the $()^\cbullet$ map as a linear form like \ref{F_linear_representation} and that the values $z_0, w_0$ are extracted from equations \ref{ext_pair_conj_axiom_Coeff_equations}.

\subsection{The pure complex case} \label{pure_complex_case}

The fulfillment of the Associativity property for the Standard Product depends on the choice of parameters $z_0, w_0$. Nothing has been imposed on the latest parameters, being these properties the constraints they must satisfy. For example, the most straight forward solution is the one applied to the complex product, where:
\begin{equation}
z_0 = 0, \; w_0 = -1,\; a_1 = -1\; a_2 = 0\; a_3 = 0\; a_4 = 1.
\end{equation}
These values lead to the linear function
\begin{equation}
\mathcal{F}_E(\gamma_E, \gamma_I) \extk + \mathcal{F}_I(\gamma_E, \gamma_I) = - \gamma_E \extk + \gamma_I,
\end{equation}
which we can recognize as the function for the complex conjugate operator $()^*$. The choice $\extk^2 = -1$ conflicts with the definition of the complex unit $\imagi^2 = -1$. Because of that, for the extended numbers, the parameter $z_0$ can not be zero.

The above equations \ref{ext_pair_conj_axiom_Coeff_equations} have multiple solutions. However, according to the main objective of this work, we need to analyze and choose the solutions that best fit the new proposal for the quantum mechanics that includes variable masses. We already show that $z_0 \neq 0$, but also it is essential to verify the solution that has a more useful meaning for future theory. The ``useful'' feature is related to the equation:
\begin{equation}
\alpha = \alpha^\cbullet, \label{number_equal_conjugate_equation}
\end{equation}
which is automatically satisfied by the inner product $\extproduct{\alpha}{\beta}{\alpha}{\beta}$, as we will show in the next sections. This quantity is related to quantum stationary states and also with fundamental concepts like measurement and the normalization condition. Indeed, the equation $z = z^*$, where $z\in \mathbb{C}$, lead to the result that $z$ is a real number. This result is the key to obtaining the stationary equation for any physical observable in Quantum Mechanics. In our case, the classical theory \cite{Israel:1811.12175} establish equations that have complex solutions. We can look for the solutions of equations \ref{ext_pair_conj_axiom_Coeff_equations}, whose values lead to pure complex numbers as the result of the equation \ref{number_equal_conjugate_equation}. 

With the linear form of equation $^\cbullet$ map, the equations \ref{number_equal_conjugate_equation} can be rewritten as:
\begin{align}
\alpha_E = a_1 \alpha_E + a_2 \alpha_I
\nonumber \\
\alpha_I = a_3 \alpha_E + a_4 \alpha_I. \label{number_equal_conjugate_equation_0}
\end{align}
If we set $a_2 = 0$ and $a_4 = 1$, the equations have the form
\begin{align}
\alpha_E &= a_1 \alpha_E
\nonumber \\
0 &= a_3 \alpha_E,
\end{align}
that lead to $\alpha_E = 0, \; \forall\; a_1\neq 1\; ||\; a_3\neq 0$, or what is the same, that all extended numbers that satisfied equation \ref{number_equal_conjugate_equation} are pure complex numbers. Replacing $a_2 = 0$ and $a_4 = 1$ on the set of equations \ref{ext_pair_conj_axiom_Coeff_equations}, we obtain:
\begin{align}
z_0 (a_1^2- a_1 ) + 2 a_1 a_3 &= 0
\nonumber \\
w_0 (a_1^2 - 1) - z_0 a_3 + a_3^2 &= 0. \label{ext_pair_conj_axiom_Coeff_equations_complex}
\end{align}

The choice to set the variables $a_2 = 0$ and $a_4 = 1$ reduce the equations \ref{number_equal_conjugate_equation} to a set of two equations with four variables $a_1, a_3, z_0,$ and $w_0$. That means we have two unconstrained variables that can have any value. The variables $z_0$ and $w_0$ are also parameters for the equations \ref{extFinalAbsRelations1}. We can propose any values for $z_0$ and $w_0$ that simplified the referred equations, as long as they satisfy $z_0 \neq 0$ and $a_1\neq 1\; ||\; a_3\neq 0$. We let this analysis and the exact computation of function $\mathcal{F}(\alpha)$ or the coefficients $a_1, a_3$ and $z_0, w_0$ for future works.

\subsection{Pair based equations for the extended inner product}
Even we define the $\mathcal{F}(\alpha)$ for the $()^\cbullet$ map, finding the functions that satisfy the properties described above, the referred map is related to the quantity $(\alpha^\bullet \odot \beta^\bullet)$ $e.i. $ with the coefficients $z_1^{(\alpha)},w_1^{(\alpha)}, z_2^{(\alpha)}, w_2^{(\alpha)}, z_1^{(\alpha^\bullet)},w_1^{(\alpha^\bullet)}, z_2^{(\alpha^\bullet)}$, and $ w_2^{(\alpha^\bullet)}$. Then, once defined the function $\mathcal{F}(\alpha)$, the coefficients must satisfy:
\begin{equation}
\mathcal{F}(\alpha \odot \beta)  = \mathcal{F}_E((\alpha \odot \beta)_E, (\alpha \odot \beta)_I) \extk + \mathcal{F}_I((\alpha \odot \beta)_E, (\alpha \odot \beta)_I)= \alpha^\bullet \odot \beta^\bullet,
\end{equation}
of explicitly:
\begin{align}
\mathcal{F}_E(\alpha, \beta, z_1^{(\alpha)},w_1^{(\alpha)}) =& \;
z_1^{(\alpha^\bullet)} z_2^{(\alpha)^*} w_2^{(\beta)} \alpha_E^* \beta_E 
+ z_1^{(\alpha^\bullet)} z_2^{(\alpha)^*} \alpha_E^* \beta_I 
+ z_2^{(\beta)} w_2^{(\alpha)^*}  \alpha_E^* \beta_E 
\nonumber \\
&+ z_2^{(\beta)} \alpha_I^* \beta_E  
\nonumber \\
\mathcal{F}_I(\alpha, \beta, z_1^{(\alpha)},w_1^{(\alpha)}) =& \;
\imagi z_2^{(\alpha)^*} z_2^{(\beta)} \alpha_E^* \beta_E 
+ z_2^{(\alpha)^*} w_1^{(\alpha^\bullet)} w_2^{(\beta)} \alpha_E^* \beta_E 
+ z_2^{(\alpha)^*} w_1^{(\alpha^\bullet)} \alpha_E^* \beta_I
\nonumber \\
&+ w_2^{(\alpha)^*} w_2^{(\beta)} \alpha_E^* \beta_E 
+ w_2^{(\alpha)^*} \alpha_E^* \beta_I
+ w_2^{(\beta)} \alpha_I^* \beta_E
+ \alpha_I^* \beta_I, \label{pair_extConjRelations_0}
\end{align}
where we replace $\mathcal{F}_i((\alpha \odot \beta)_E, (\alpha \odot \beta)_I) \equiv \mathcal{F}_i(\alpha, \beta, z_1^{(\alpha)},w_1^{(\alpha)})$ $\forall\; i=E,I$, to show the $z_1, z_2, w_1, w_2$ dependency of the arguments.

From the inner product chapter \ref{inner_prod_chapter}, and using the Positive definiteness axiom of the number $\alpha$, we were able to identify only six complex equations from the eight needed for computing the eight later coefficients $z_1^{(\alpha)},w_1^{(\alpha)}, z_2^{(\alpha)}, w_2^{(\alpha)}, z_1^{(\alpha^\bullet)},w_1^{(\alpha^\bullet)}, z_2^{(\alpha^\bullet)}$ and $ w_2^{(\alpha^\bullet)}$. We need one extended or two complex equations more for fully compute the coefficients. On the other side, the equation set \ref{pair_extConjRelations_0} involves the coefficient of the extended numbers $\alpha$ and $\beta$. That means that we also need to find the unknown coefficients $z_1^{(\beta)},w_1^{(\beta)}, z_2^{(\beta)}, w_2^{(\beta)}, z_1^{(\beta^\bullet)},w_1^{(\beta^\bullet)}, z_2^{(\beta^\bullet)}$ and $ w_2^{(\beta^\bullet)}$ of the extended number $\beta$. Adding its respective coefficients and the set of equations \ref{extConjRelations_1} for the number $\beta$ have 16 coefficients to determine and two sets of equations \ref{extConjRelations_1} for a total of 12 equations. The previous equations \ref{pair_extConjRelations_0} can be added to the full set of equations letting only one extended equation to find. Also, since $\alpha$ and $\beta$ are two aleatory numbers, they should satisfy the same equations. As equations \ref{pair_extConjRelations_0} are not symmetric for those numbers, we need to include the same equations \ref{pair_extConjRelations_0}, this time for the expression $(\beta \cdot \alpha)$:
\begin{equation}
\mathcal{F}(\beta\odot \alpha ) = \mathcal{F}_E((\beta \odot \alpha)_E, (\beta \odot \alpha)_I) \extk + \mathcal{F}_I((\beta \odot \alpha)_E, (\beta \odot \alpha)_I)= \beta^\bullet \odot \alpha^\bullet,
\end{equation}
of explicitly:
\begin{align}
\mathcal{F}_E(\beta, \alpha, z_1^{(\beta)},w_1^{(\beta)}) =& \;
z_1^{(\beta^\bullet)} z_2^{(\beta)^*} w_2^{(\alpha)} \beta_E^* \alpha_E 
+ z_1^{(\beta^\bullet)} z_2^{(\beta)^*} \beta_E^* \alpha_I 
+ z_2^{(\alpha)} w_2^{(\beta)^*}  \beta_E^* \alpha_E 
\nonumber \\
&+ z_2^{(\alpha)} \beta_I^* \alpha_E  
\nonumber \\
\mathcal{F}_I(\beta, \alpha, z_1^{(\beta)},w_1^{(\beta)}) =& \;
\imagi z_2^{(\beta)^*} z_2^{(\alpha)} \beta_E^* \alpha_E 
+ z_2^{(\beta)^*} w_1^{(\beta^\bullet)} w_2^{(\alpha)} \beta_E^* \alpha_E 
+ z_2^{(\beta)^*} w_1^{(\beta^\bullet)} \beta_E^* \alpha_I
\nonumber \\
&+ w_2^{(\beta)^*} w_2^{(\alpha)} \beta_E^* \alpha_E 
+ w_2^{(\beta)^*} \beta_E^* \alpha_I
+ w_2^{(\alpha)} \beta_I^* \alpha_E
+ \beta_I^* \alpha_I, \label{pair_extConjRelations_1}
\end{align}
\subsection{Extended Pair Conjugate symmetry}
Once the Associativity for the Sum and the Standard Product is satisfied by the operator, $()^\cbullet$, we can show that the inner product satisfies the extended Pair Conjugate symmetry. Indeed, the Extended Pair Conjugate of the extended product of four extended numbers:
\begin{equation}
\extproduct{\alpha}{\beta}{\gamma}{\delta}^\cbullet \equiv [(\alpha^\bullet \odot \beta^\bullet)\cdot(\gamma \odot \delta)]^\cbullet = (\alpha^\bullet \odot \beta^\bullet)^\cbullet \cdot (\gamma \odot \delta)^\cbullet,
\end{equation}
applying the Associative property of the standard product. On the other side, the application of the map 
\begin{equation}
(\alpha^\bullet \odot \beta^\bullet)^\cbullet \cdot (\gamma \odot \delta)^\cbullet
= ((\alpha^\bullet)^\bullet \odot (\beta^\bullet)^\bullet)\cdot(\gamma^\bullet \odot \delta^\bullet),
\end{equation}
can be modified using the closure condition \ref{closureCondition_0}, which state that $(\alpha^\bullet)^\bullet = \alpha$ and obtain:
\begin{equation}
 ((\alpha^\bullet)^\bullet \odot (\beta^\bullet)^\bullet)\cdot(\gamma^\bullet \odot \delta^\bullet)
 = (\alpha \odot \beta)\cdot(\gamma^\bullet \odot \delta^\bullet) \equiv \extproduct{\gamma}{\delta}{\alpha}{\beta},
\end{equation}
proving our initial statement and providing the space with a bi-linear form.

\section{Final equations to determine coefficients $z_i, w_i$}
Along the develop of this work we has obtained some sets of equations for computing the coefficients  $z_1^{(\alpha)},w_1^{(\alpha)}, z_2^{(\alpha)}, w_2^{(\alpha)}$ related to the unknown complex product $()\odot()$ and the map $()^\bullet$. Some of the wanted equations has arrive from the axioms that the extended inner products must satisfy, according our main objective. Also, some of the parameter included on such equations are proposed from logical analysis. This section resumes all the equations and states the method for computing the wanted coefficients.

From the positive-definiteness, we obtain a set of six complex equations, equations \ref{extFinalAbsRelations1} and \ref{closureCondition}, relating the coefficients $z_1,w_1, z_2, w_2$ for the extended numbers $\alpha$ and $\alpha^\bullet$. We represent this set of equation as 
\begin{equation}
\mathcal{D}_n(\alpha, \alpha^\bullet)=0,\;  \forall\; 0 \leq n \leq 6 \label{D_equations}
\end{equation}
These equations, which are insufficient for fully determine the wanted coefficients, depends on the parameters $z_0$, $w_0,$ and $R$. The $R$-parameter was chosen to have value $R=2$ and the $z_0$, $w_0,$ are determined by the equations \ref{ext_pair_conj_axiom_Coeff_equations} proposed from the properties of the $()^\cbullet$ map. Also, the map $()^\cbullet$ is related with the complex product $\alpha \odot \beta$ and  $\beta \odot \alpha$ as showed in the previous section.

For obtaining the coefficients $z_i, w_i$ of any extended number, the general method is: first obtaining the $z_0$, $w_0,$ parameters from the equations \ref{ext_pair_conj_axiom_Coeff_equations_complex}. That means that the linear form of the function $\mathcal{F}$, equation \ref{F_linear_representation} for the $()^\cbullet$ map, is fully determined. We can then, impose the equations \ref{pair_extConjRelations_0} and \ref{pair_extConjRelations_1}. As these equations are related to the complex product of two extended numbers, we need to add the equations \ref{D_equations} for both numbers. The set of complex equations from equations \ref{D_equations}, \ref{pair_extConjRelations_0}, and \ref{pair_extConjRelations_1} for the described situation is
\begin{align}
\mathcal{D}_n(\alpha, \alpha^\bullet)&=0
\nonumber \\
\mathcal{D}_n(\beta, \beta^\bullet)&=0
\nonumber \\
 (\alpha^\bullet \odot \beta^\bullet)_E &= \mathcal{F}_E((\alpha \odot \beta)_E, (\alpha \odot \beta)_I)
\nonumber \\
(\alpha^\bullet \odot \beta^\bullet)_I &= \mathcal{F}_I((\alpha \odot \beta)_E, (\alpha \odot \beta)_I)
\nonumber \\
 (\beta^\bullet \odot \alpha^\bullet)_E &= \mathcal{F}_E((\beta \odot \alpha)_E, (\beta \odot \alpha)_I)
\nonumber \\
(\beta^\bullet \odot \alpha^\bullet)_I &= \mathcal{F}_I((\beta \odot \alpha)_E, (\beta \odot \alpha)_I), \label{final_equations0}
\end{align}
summing the 16 equations needed for determining the 16 coefficients $z_i, w_i$ of numbers $\alpha, \beta, \alpha^\bullet, \beta^\bullet$. 

However, following the main goal of this work, we set some values for the coefficients of the linear  expression of the $()^\cbullet$ map, as seen in section \ref{pure_complex_case}. The choice of the values, ensures that the inner product
\begin{equation*}
\extproduct{\alpha}{\beta}{\alpha}{\beta} \quad \in \; \mathbb{C}.
\end{equation*}
This lead to two unconstrained coefficients that can have any values with some exceptions. In this case, we can follow other method and first check the $\mathcal{D}_n(\alpha, \alpha^\bullet)$ and propose values for $z_0$ and $w_0$ that simplified those equations, then compute the others coefficients $a_1$ and $a_3$, finding the final form of function $\mathcal{F}$. We set then the final equations \ref{final_equations0} for computing the coefficients $z_i, w_i$ for numbers $\alpha, \alpha^\bullet, \beta$ and $\beta^\bullet$. Replacing the linear representation of function $\mathcal{F}$, equation \ref{F_linear_representation}, the final set of equations is

\begin{align}
\mathcal{D}_n(\alpha, \alpha^\bullet)&=0
\nonumber \\
\mathcal{D}_n(\beta, \beta^\bullet)&=0
\nonumber \\
(\alpha^\bullet \odot \beta^\bullet)_E &= a_1(\alpha \odot \beta)_E 
\nonumber \\
(\alpha^\bullet \odot \beta^\bullet)_I &= a_3(\alpha \odot \beta)_E + (\alpha \odot \beta)_I
\nonumber \\
 (\beta^\bullet \odot \alpha^\bullet)_E &=a_1(\beta \odot \alpha)_E
\nonumber \\
 (\beta^\bullet \odot \alpha^\bullet)_I&= a_3(\beta \odot \alpha)_E + (\beta \odot \alpha)_I, \label{final_equations}
\end{align}
where coefficients $a_1, a_3$ satisfy equations \ref{ext_pair_conj_axiom_Coeff_equations_complex}
\begin{align*}
z_0 (a_1^2- a_1 ) + 2 a_1 a_3 &= 0
\nonumber \\
w_0 (a_1^2 - 1) - z_0 a_3 + a_3^2 &= 0,
\end{align*}
and also $z_0 \neq 0$ and $a_1\neq 1\; ||\; a_3\neq 0$.

The solution of these equations provide a two components vector linear space with a bi-dimensional form. The proposed equations are numerous and have a high complexity. We let this problem to be solved on future works.
\section{Conclusions}
In this work, we proposed a new vector space that includes negative probabilities. A new domain of numbers is proposed from the definition of a new algebraic unit from an unsolvable equation in the complex domain. We identified that the new set of numbers defined with such a unit includes two new types of operations for not violate the fundamental theorem of algebra. The operations include the sum and the multiplication of conjugated numbers, whose form was expressed as coefficients. Because of the positive-definiteness, we define a new map and also expressed with unknown coefficients. The vector space is provided with the positive-definiteness, pair-pair linearity, and a pair-pair conjugated symmetry that provided the independent set of equations needed for computing the coefficients from the new operations. That is our starting point, which we try to apply it to the already known algebraic concepts. The present proposal is just a preliminary study that should be enhanced.

\bibliography{References}

\end{document}